\documentclass[twocolumn]{pasj01}

\Received{2021/04/07}
\Accepted{2021/12/25}
\usepackage{url}
\usepackage{siunitx}
\newcommand{\red}[1]{{#1}}
\newcommand{\redd}[1]{{#1}}
\newcommand{\reddd}[1]{{#1}}
\newcommand{\redddd}[1]{{#1}}

\begin{document}

\title{ 
Simulation-based spectral analysis of X-ray CCD data affected by photon pile-up}

\author{Tsubasa \textsc{Tamba}\altaffilmark{1,}${^{*}}$}
\author{Hirokazu \textsc{Odaka}\altaffilmark{1,2,3}}
\author{Aya \textsc{Bamba}\altaffilmark{1,3}}
\author{Hiroshi \textsc{Murakami}\altaffilmark{4}}
\author{Koji \textsc{Mori}\altaffilmark{5,9}}
\author{Kiyoshi \textsc{Hayashida}\altaffilmark{6,7,9}}
\author{Yukikatsu \textsc{Terada}\altaffilmark{8,9}}
\author{Tsunefumi \textsc{Mizuno}\altaffilmark{10}}
\author{Masayoshi \textsc{Nobukawa}\altaffilmark{11}}
\altaffiltext{1}{Department of Physics, The University of Tokyo, 7-3-1 Hongo, Bunkyo-ku, Tokyo 113-0033, Japan}
\altaffiltext{2}{Kavli IPMU (WPI), UTIAS, The University of Tokyo, 5-1-5 Kashiwanoha, Chiba 277-8583, Japan}
\altaffiltext{3}{Research Center for the Early Universe, School of Science, The University of Tokyo, 7-3-1 Hongo, Bunkyo-ku, Tokyo 113-0033, Japan}
\altaffiltext{4}{Faculty of Liberal Arts, Tohoku Gakuin University, 2-1-1 Tenjinzawa, Izumi-ku, Sendai, Miyagi 981-3193, Japan}
\altaffiltext{5}{Department of Applied Physics and Electronic Engineering, University of Miyazaki, 1-1 Gakuen Kibana-dai Nishi, Miyazaki 889-2192, Japan}
\altaffiltext{6}{Department of Earth and Space Science, Osaka University, 1-1 Machikaneyama-cho, Toyonaka, Osaka 560-0043, Japan}
\altaffiltext{7}{Project Research Center for Fundamental Sciences, Osaka University, 1-1 Machikaneyama-cho, Toyonaka, Osaka 560-0043, Japan}
\altaffiltext{8}{Graduate School of Science and Engineering, Saitama University, 255 Shimo-Ohkubo, Sakura, Saitama 338-8570, Japan}
\altaffiltext{9}{Japan Aerospace Exploration Agency, Institute of Space and Astronautical Science, 3-1-1 Yoshino-dai, Chuo-ku, Sagamihara, Kanagawa
252-5210, Japan}
\altaffiltext{10}{Hiroshima Astrophysical Science Center, Hiroshima University, 1-3-1 Kagamiyama, Higashi-Hiroshima, Hiroshima, 739-8526, Japan}
\altaffiltext{11}{Faculty of Education, Nara University of Education, Takabatake-cho, Nara, Nara 630-8528, Japan}
\email{tsubasa.tamba@phys.s.u-tokyo.ac.jp}



\maketitle

\begin{abstract}
We have developed a simulation-based method of spectral analysis for pile-up affected data of X-ray CCDs without any loss of photon statistics. 
As effects of the photon pile-up appear as complicated nonlinear detector responses, we employ a detailed simulation to calculate the important processes in an X-ray observation including physical interactions, detector signal generation, detector readout, and a series of data reduction processes. 
This simulation naturally reproduces X-ray-like and background-like events as results of X-ray photon merging in a single pixel or in a chunk of adjacent pixels, allowing us to construct a nonlinear spectral analysis framework that can treat pile-up affected observation data. 
For validation, we have performed data analysis of \textit{Suzaku} XIS observations by using this framework with various parameters of the detector simulation all of which are optimized for that instrument. 
We present three cases of different pile-up degrees: PKS~2155-304 (negligible pile-up), Aquila~X-1 (moderate pile-up), and the Crab Nebula (strong pile-up); we show that the nonlinear analysis method produces results consistent with a conventional linear analysis for the negligible pile-up condition, and accurately corrects well-known pile-up effects such as spectral hardening and flux decrease for the pile-up cases. 
These corrected results are consistent with those obtained by a widely used core-exclusion method or by other observatories with much higher timing resolutions (without pile-up). 
Our framework is applicable to any types of CCDs used for X-ray astronomy including a future mission such as \textit{XRISM} by appropriate optimization of the simulation parameters. 
\end{abstract}


\section{Introduction}
The charge-coupled device (CCD) has been serving as a standard detector in soft X-ray imaging and spectroscopy since its first space operations with the SIS \citep{Burke1991} onboard \textit{ASCA} \citep{Tanaka1994}. With its good energy and spatial resolutions, it has been adopted by many X-ray astronomy missions such as \textit{Chandra} \citep{Weisskopf2002}, \textit{XMM-Newton} \citep{Jansen2001}, \textit{Swift} \citep{Gehrels2004}, \textit{Suzaku} \citep{Mitsuda2007}, MAXI \citep{Matsuoka2009}, and \textit{Hitomi} \citep{Takahashi2016}. \textit{XRISM} \citep{Tashiro2018}, scheduled for launch in 2022, will also utilize CCDs named Xtend-SXI \citep{Hayashida2018} together with a micro-calorimeter \textit{Resolve} \red{\citep{Ishisaki2018}}, aiming at imaging and spectroscopy in the soft X-ray band.

Since the CCD needs a readout time of a few seconds, photon pile-up has always been a serious problem when observing bright sources. When multiple photons fall onto the same or nearby pixels within a single frame readout, these photons are merged into one and analyzed as a single event with an energy equivalent to the summation of the incident photon energies. This merged event is processed as an X-ray event or excluded as a noise-like event, both of which lead to a distortion of the measured spectrum of photon counts. Recent observatories have experienced pile-up problems more seriously due to the improvement of the photon collection capability. In the case of XRISM-Xtend, for example, the large effective area and the sharp point spread function (PSF) of the X-ray mirror, the thick depletion layers of the CCDs ($\sim200\;{\rm \mu m}$), and the relatively large pixel size of the CCDs ($48\;{\rm \mu m}\times48\;{\rm \mu m}$ generated by binning $2\times2$ pixels into 1 pixel) all contribute to increase the average number of photons per pixel. Based on an observation with \textit{Hitomi}, which employed almost the same CCDs and X-ray mirrors as those of \textit{XRISM}, sources at the level of $\sim1\;{\rm mCrab}$ could be affected by pile-up (e.g., NGC~1275: \cite{HitomiCollaboration2018NGC1275}). \red{In order to make use of the micro-calorimeter onboard \textit{XRISM}, many targets will be brighter than this flux level.} Therefore, it \red{is} important to develop a method to treat pile-up affected observation data.

The major effects on detector count spectra caused by the pile-up consist of hardening of the spectrum and decrease in count rates, resulting in a harder photon index and a lower flux than their true values. The distortion of the  spectrum is a complicated process depending on a spectral index, a normalization, a mirror effective area, the quantum efficiency of a detector, and many other properties. Due to these complexities, it is difficult to derive the correct spectrum from irreversible pile-up affected data.

A commonly-used technique to eliminate the pile-up effects is the ``core-exclusion'' method.
This only analyzes events in low-count-rate regions, excluding the pile-up affected image center. \citet{Yamada2012} performed further investigations on this method, and developed criteria which determine the minimum radius of the core-exclusion by using a measurable indicator of the pile-up degree called ``pile-up fraction''.
Although this approach was effective to some extent and contributed to many studies, it produces large statistical errors due to a huge loss of the photon counts. Some previous studies have proposed other methods. \citet{Davis2001} successfully constructed a spectral analysis model by attributing the pile-up effects to a newly introduced parameter of ``grade migration'' probability. 
However, this method does not take account of the energy dependence of the pile-up degree, which is caused by the differences by energy in the distributions of charge clouds and PSF.
\red{\citet{Ballet1999} and \citet{Ballet2003} present another theoretical study for CCD pile-up for detectors with small pixel sizes compared to the PSF by focusing on dominant single-pixel events, but it cannot be applied to detectors generating many multi-pixel events such as back-illuminated CCDs.}
In order to construct an effective way to treat pile-up affected data without the loss of the photon statistics, it is necessary to correctly treat the effects of the pile-up on the detector response. A Monte Carlo simulation is an effective way to treat the complicated processes related to the pile-up.

In this paper, we present a new framework for spectral analysis of pile-up affected data by using Monte Carlo simulation, and carry out a comprehensive study on how detector count spectra would be distorted by the pile-up. We treated \textit{Suzaku} XIS data to validate our method applied to the pile-up affected data in real observations. The basic concepts and designs of the simulator are presented in section \ref{section:development_of_CCD_simulator}. After tuning several simulation parameters in section \ref{section:reproduction_of_observation}, we evaluated the spectral distortion in section \ref{section:model_simulation}. The method and results of spectral analysis on \textit{Suzaku} XIS data using our framework are described in section \ref{section:spectral_analysis_for_pile-up_data} . The discussion is presented in section \ref{section:discussion} and the conclusions are given in section \ref{section:conclusions}.

\section{Simulation model of CCD responses}\label{section:development_of_CCD_simulator}
In the standard procedure of the X-ray spectral analysis for a point source, one usually employs the relation between the incident source spectrum $S(E)$ and the \reddd{detector} count spectrum $C(h)$,
\begin{eqnarray}
C(h)=\int R(E, h)A(E)S(E)dE,
\label{eq:count_spectrum}
\end{eqnarray}
where $R(E, h)$ and $A(E)$ denote the response matrix and the effective area of the detector, respectively (also see \cite{Arnaud1996} for {\tt XSPEC}, one of the most commonly-used spectral analysis tools). This relation between the space of incident photon energies $E$ and the space of pulse heights $h$ is assumed to be linear, allowing us to apply a spectral fitting using a linear response matrix with chi-square statistics or c-statistics. However, the pile-up breaks the linearity of the response matrix by merging multiple photon events. \red{Then, the response function becomes a nonlinear form, which depends on the incident spectral shape and effective area function. \reddd{As a result, one observes a distorted spectrum $C'(h)$ rather than the intrinsic $C(h)$. Hereafter in this paper, we denote $C(h)$ as ``intrinsic count spectrum'' and $C'(h)$ as ``observed count spectrum'' to distinguish them.} The \reddd{observed (distorted)} count spectrum $C'(h)$ is hard to obtain since one needs to calculate the nonlinear response function in every step of spectral fitting.} Moreover, the additional information needed for an analytical approach such as \red{event shape (Grade) distribution (generally obtained for only several energies), grade migration, and merging of sub-threshold residual charges caused by pile-up is hard to obtain}, which makes the analytical approach less efficient (for a detailed analytical approach, see \cite{Davis2001}).
Instead of the analytical calculation, we adopt a Monte Carlo simulation to obtain $C'(h)$, which makes it easy to treat nonlinear pile-up effects. In this section, we present the basic concept and structure of the simulation.

\begin{figure*}
\begin{center}
\includegraphics[width=160mm]{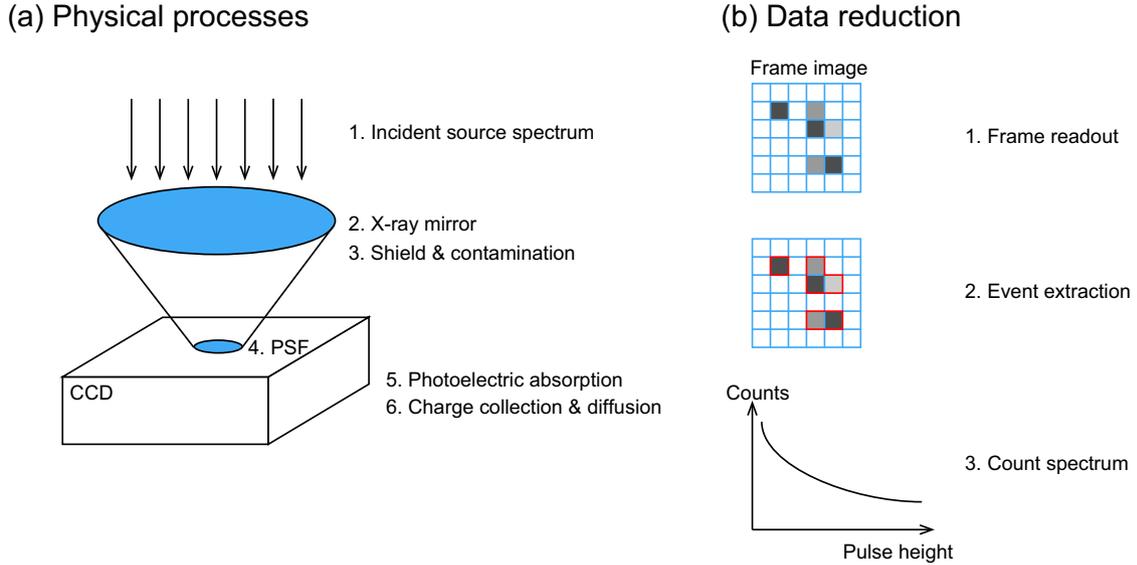}
\end{center}
\caption{The processes in an X-ray CCD observation that the simulator takes into account. See text in detail.}
\label{fig:simulator_basic_concepts}
\end{figure*}

Our simulator handles the nonlinear detector response by calculating all the processes in detecting source photons. It includes not only physical processes incident photons experience and induce but also data acquisition and reduction processes which convert electrical signals to detection information. The details of the two processes are described in the following subsections. An overview of the entire processes is also presented in Figure \ref{fig:simulator_basic_concepts}.

We implemented all the functions of the simulator in ComptonSoft \citep{Odaka2010}, which is a framework for X-ray detector simulation and data analysis. In ComptonSoft, Monte Carlo simulation of photon interactions with detectors is based on Geant4 \citep{Agostinelli2003, Allison2006, Allison2016}, which is a library of Monte Carlo simulation for high energy particle physics. 
All the simulations presented in this paper were conducted using ComptonSoft 5.7.0 and Geant4 10.05.01.

\subsection{Physical processes---simulation of photon detection}\label{section:physics}
The first step of the simulator calculates physical processes of incident photon interactions and signal generation in detectors. All the processes are presented in the left panel of Figure \ref{fig:simulator_basic_concepts}. The input of this step consists of the incident source spectrum $S(E)$, the effective area function $A(E)$, the point spread function (PSF), and the observation time. We describe here the details of each step.
\begin{enumerate}
{\bf \item Photon incidence}\\
The simulator accepts any shape (e.g., a power law, a black body, or a user-defined histogram) of the incident source spectrum $S(E)$ in units of ${\rm photons\;s^{-1}\;cm^{-2}\;keV^{-1}}$.
{\bf \item Photon collection by the X-ray mirror}\\
The number of photons collected is determined by the effective area of the X-ray mirror. It is expressed as $A(E)$, a function of incident photon energy. The simulator reads the effective area from the ancillary response file (ARF).
{\bf \item Transmission of the shield and contamination}\\
The incident photon is attenuated by the shield and contamination before it reaches the detector surface. The transmission rates of these materials are also read from the ARF, if the corresponding columns exist in the table. Therefore, the input spectrum used in the Monte Carlo simulation is determined as the photon spectrum immediately above the detector surface. This is given by
\begin{eqnarray}
S_{\rm input}(E) = S(E)A(E) T_{\rm shild}(E) T_{\rm contam}(E)
\label{eq:effective_area_all}
\end{eqnarray}
in units of ${\rm photons\;s^{-1}\;keV^{-1}}$, where $T_{\rm shield}(E)$, and $T_{\rm contam}(E)$ represent the transmission rate of the shield and that of the contamination, respectively.
{\bf \item Spatial distribution of incident photons}\\
The PSF denotes horizontal distribution of photons from a point source at the detector surface. The simulator reads the PSF, and assigns one incident position to each incident photon by sampling a position from the spatial distribution described by the PSF. 
{\bf \item Photon interactions with the detector}\\
Photons arriving at the detector either interact with the detector material or transmit through it. Most of the interactions are due to photoelectric absorption which is followed by ejection of secondary electrons or photons. A series of these processes is simulated by Geant4 with a detector geometry description. For each incident event, the simulator extracts information on the incident photon and all the subsequent energy deposits, from which we obtain properties of the initial charge cloud generated in the detector.
{\bf \item Charge cloud collection and diffusion}\\
The charge cloud generated by the photoelectric absorption is collected by an applied electric field, experiencing diffusion while drifting to electrodes. The drift and diffusion effects are calculated in the simulator so that one can obtain how many charges are collected to each electrode, yielding the pulse heights of all the pixels relevant to each incident photon.\\
The number of electron-hole pairs associated with the initial charge cloud is fluctuated via Poisson distribution and the readout signal also has electronics noise including dark current. The dependence of noise on energy can vary with its physical origin. In the simulator, we assume the energy resolution is composed of three terms as
\begin{eqnarray}
\Delta E = \sqrt{p_{0}{}^2+\left(p_{1}\sqrt{E}\right)^2+\left(p_2E\right)^2},
\label{eq:energy_resolution}
\end{eqnarray}
each term representing contribution from $E^0$, $E^{1/2}$, and $E$, respectively. The coefficients $p_0$, $p_1$, and $p_2$ are given by users as simulation parameters.
\end{enumerate}

The output of this part---the simulation of the physical processes---is a list of events each of which contains an initial photon energy and electric signals on detector pixels generated by all energy deposits due to the physical interactions in the detector as described above. Here, an event corresponds to an incidence of a photon. \redd{Any finite charge (non-zero) produced by electron ionization is recorded in the simulation.}

\subsection{Data readout---frame readout and event extraction}
The second step of the simulator treats data reduction processes. Three steps in this part are presented in the right panel of Figure \ref{fig:simulator_basic_concepts}. Here we describe detailed implementations of those steps. This part takes an output of the physics part (Section \ref{section:physics}) as an input. The frame exposure time and event and split threshold energies for the event extraction are set by users as simulation parameters.

\begin{enumerate}
{\bf \item Frame readout}\\
A frame image presents how many charges are collected at each pixel within a specific exposure time. In real observations with a CCD, this process is carried out on-line in orbit. If the pile-up effects are negligible, the pulse height of each pixel in each frame simply represents all or a part of the charges generated by a single incident photon. If the pile-up effects exist, on the other hand, it could be the summation of the charges generated by overlapping multiple photons. We simulate this merging process of multiple photons within a single frame exposure by assigning a randomized arrival time to each incident photon.

{\bf \item Event extraction}\\
Event extraction from a frame image is done by the so-called Grade algorithm. It extracts event regions by using event and split thresholds, classifies their event shapes into several patterns, and determines whether these events are caused by an incident X-ray or not. We implemented the Grade algorithm which is employed by \textit{Suzaku} XIS \citep{Koyama2007}, \textit{Hitomi} SXI \citep{Tsunemi2010}, and \textit{XRISM} Xtend-SXI \citep{Hayashida2018}. \red{What should be noted here is that the pile-up not only merges multiple events but also generates new events which are detected due to summation of multiple sub-threshold signals. Our simulation model can reproduce both of these effects in the event extraction process.}

{\bf \item Generation of \reddd{observed} count spectrum}\\
Finally, the simulator fills a histogram with the energies of the extracted events, generating an observed count spectrum. At this stage, we obtain the \reddd{observed} count spectrum distorted by the pile-up effects ($C'(h)$) rather than the \reddd{intrinsic} count spectrum without any pile-up effects ($C(h)$, equation (\ref{eq:count_spectrum})). The obtained spectrum can be used for spectral analysis by comparing it with observation data.

\end{enumerate}


\section{Simulations applied to \textit{Suzaku} XIS observations}\label{section:reproduction_of_observation}
In order to apply our simulator to a specific detector, it is necessary to optimize various simulation parameters by comparing the simulation outputs with officially provided detector responses or real observation data. It leads to ensuring that the simulator accurately reproduces real observations. In this work, we used observation data with \textit{Suzaku} \citep{Mitsuda2007}, which was in operations from 2005 to 2015. The \textit{Suzaku} was equipped with four CCDs (XIS0 through 3) \citep{Koyama2007}, which are classified into two types; XIS0, 2 and 3 are front-illuminated (FI) CCDs while XIS1 is a back-illuminated (BI) CCD. Since the FI and BI have totally different properties, we performed parameter tuning for each detector individually.

In this section and the following sections, we utilized {\tt XSPEC} 12.11.0 \citep{Arnaud1996} for spectral analyses, with solar metallicity abundance model angr \citep{Anders1989} and photoelectric absorption cross-section model vern \citep{Verner1996}. We used {\tt aepipeline} in {\tt HEASoft} 6.27.2 for data reduction. The generations of response matrices and ancillary responses were performed by {\tt xisrmfgen} and {\tt xissimarfgen} respectively, referring to CALDB (the HEASARC Calibration Database). \redd{The PSF is read from the ARF file. We did not sample the PSF at the sub-pixel level because the size of the PSF is much larger than the pixel size in the case of Suzaku.}

Table \ref{tab:simulation_parameter} shows a list of simulation parameters not necessary to tune as they are explicitly presented in previous researches or in official information. They include the geometry of the CCDs and properties of surrounding substances. In addition to these parameters, we have to optimize several parameters so that the simulator can accurately reproduce real observations. These parameters to be optimized are the geometry of the dead region, the depletion layer thickness, and the electric field structure. Although the depletion layer thickness is presented in the previous study \citep{Koyama2007}, we still need to optimize it to duplicate the detector response accurately. Note that the goal of the parameter tuning is producing an accurate detector response by the simulation, and the best-fit values of the parameters do not necessarily reflect the accurate physical properties due to parameter coupling (degeneracy).

\begin{table*}
  \tbl{List of simulation parameters\footnotemark[$*$] }{%
  \begin{tabular}{cccc}
      \hline
       & XIS0, 3 (FI) & XIS1 (BI) & References\footnotemark[$\dag$]\\ 
      \hline
      Material & \multicolumn{2}{c}{Si ($2.33\;{\rm g\;cm^{-3}}$)} & (1)\\
      Pixel size & \multicolumn{2}{c}{$24\;{\rm \mu m}\times24\;{\rm \mu m}$} & (1)\\
      Format & \multicolumn{2}{c}{$1024\times1024$ pixels} & (1)\\
      Temperature & \multicolumn{2}{c}{$-90{\rm \degree C}$} & (1)\\
      Optical blocking filter & \multicolumn{2}{c}{Al $0.12\;{\rm \mu m}$} & (1)\\
      Energy resolution\footnotemark[$\ddag$] & \multicolumn{2}{c}{$p_0=0.0215\;{\rm keV}$, $p_1=0.0209\;{\rm keV^{1/2}}$, $p_2=0$} & (1)\\
      Event threshold & \multicolumn{2}{c}{$365\;{\rm eV}$} & (2)\\
      Split threshold & $365\;{\rm eV}$ & $73\;{\rm eV}$ & (2)\\
      \hline
    \end{tabular}}\label{tab:simulation_parameter}
\begin{tabnote}
\footnotemark[$*$] For parameters which need tuning, see table \ref{tab:parameter_tuning_result}.  \\ 
\footnotemark[$\dag$] (1) \cite{Koyama2007}; (2) DARTS Astrophysics XIS config parameters  \url{http://darts.isas.jaxa.jp/astro/suzaku/data/xisconf_list.html}\\
\footnotemark[$\ddag$]  $p_1$ is determined by setting the Fano factor of Si to $0.12$. $p_0$ is set to satisfy an FWHM of $130\;{\rm eV}$ at $5.9\;{\rm keV}$ \citep{Koyama2007} with equation (\ref{eq:energy_resolution}).\\ 
\end{tabnote}
\end{table*}

\begin{table}
  \tbl{Results of simulation parameter tuning\footnotemark[$*$] }{%
  \begin{tabular}{cccc}
      \hline
       & XIS0 & XIS1 & XIS3\\ 
      \hline
      ${\rm SiO_{2}}$ dead region\footnotemark[$\dag$] & $0.6\;{\rm \mu m}$ & $0.05\;{\rm \mu m}$ & $0.6\;{\rm \mu m}$\\
      Si dead region\footnotemark[$\dag$] & $0.4\;{\rm \mu m}$ & $0.0\;{\rm \mu m}$ & $0.4\;{\rm \mu m}$ \\
      Depletion layer\footnotemark[$\ddag$] & $70\;{\rm \mu m}$ & $42\;{\rm \mu m}$ & $74\;{\rm \mu m}$\\
      Bias voltage\footnotemark[$\S$] & $19.0\;{\rm V}$ & $9.0\;{\rm V}$ & $20.4\;{\rm V}$\\ 
      \hline
    \end{tabular}}\label{tab:parameter_tuning_result}
\begin{tabnote}
\footnotemark[$*$] The table presents the best-fit parameters of tuning. The presented values do not necessarily reflect the accurate physical properties.\\ 
\footnotemark[$\dag$] The errors are $\pm0.05\;{\rm \mu m}$ for XIS0 and XIS3, $\pm0.025\;{\rm \mu m}$ for XIS1.\\
\footnotemark[$\ddag$] The errors are $\pm 1\;{\rm \mu m}$ for all the detectors.\\ 
\footnotemark[$\S$] The errors are $\pm0.1\;{\rm V}$ for all the detectors.\\ 
\end{tabnote}
\end{table}

\subsection{Parameter tuning with CALDB}\label{section:parameter_tuning_1}
The first step of the simulation parameter tuning aims at producing a detector response that reasonably agrees with the detector response provided by the official CALDB (hereafter, the CALDB response). This part was performed without using real observation data since the CALDB provides us with sufficiently accurate reference. We assumed no pile-up at this stage, as the official CALDB response does not take the pile-up effects into account. We thus set the frame exposure to a sufficiently short time, virtually $0\;{\rm s}$, so that one frame is assigned to exactly one event and pile-up never occurs.

The targets of the parameter tuning with respect to the CALDB response consist of the geometry of the dead region and the depletion layer thickness. The former mainly affects the quantum efficiencies of low energy photons while the latter mainly affects those of high energy photons. We performed simulations for various parameter sets and evaluated $\chi^2$ between a count spectrum generated through the simulation and that through the CALDB response to determine the best-fit parameter sets. To make these spectra, an incident source was assumed to have a flat spectrum $S(E)=S_0$ (constant).

\begin{figure}
\begin{center}
\includegraphics[width=80mm]{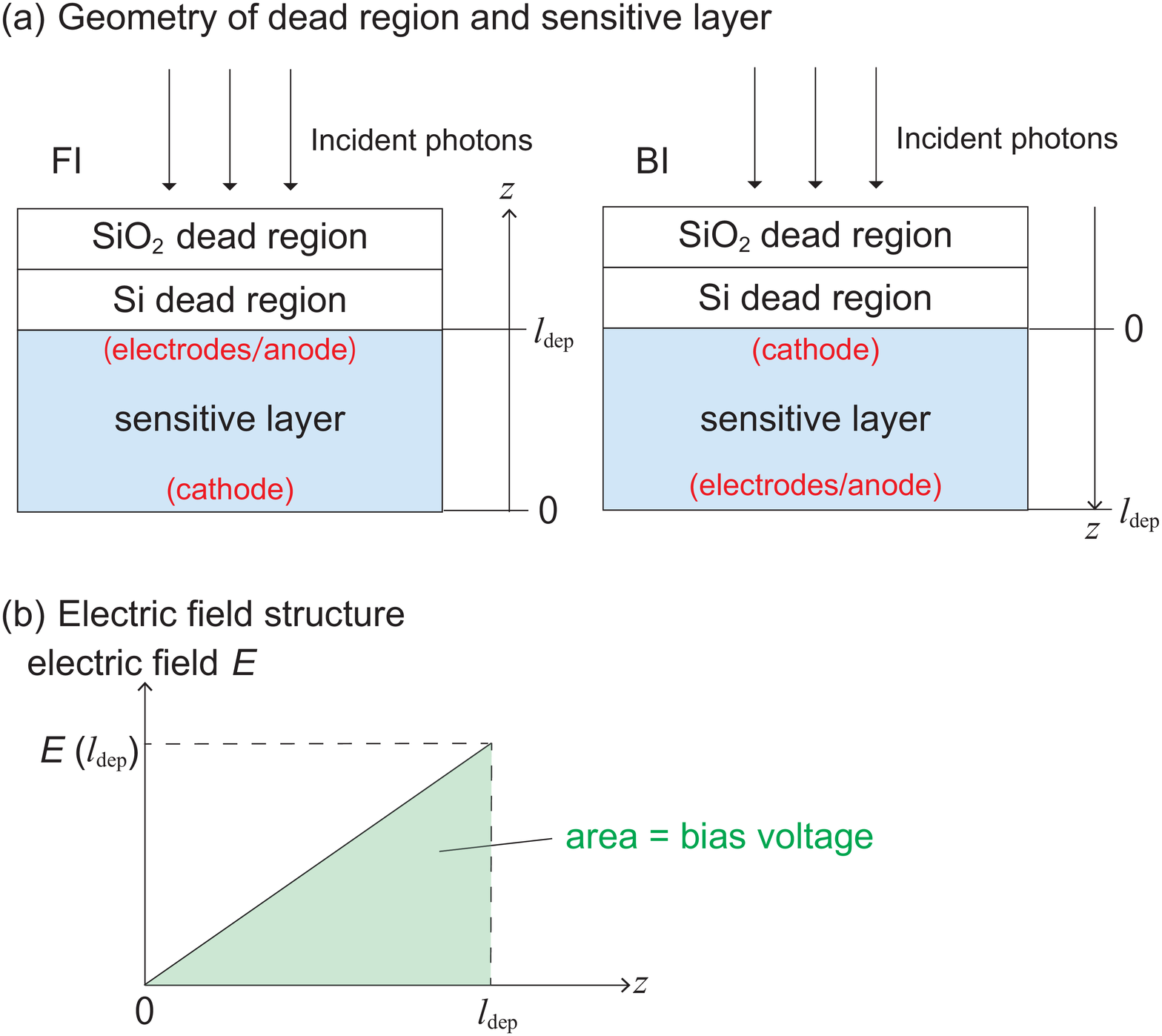}
\end{center}
\caption{(a) The geometry of our simulation setup. The dead regions lie above the sensitive layers. The $l_{\rm dep}$ is the thickness of the sensitive layer. (b) Electric field structure of the depletion layer we assumed in the simulation. The shaded area corresponds to the bias voltage.}
\label{fig:parameter_tuning_explanation}
\end{figure}

The dead region is a thin layer which lies just above the sensitive region and mainly intercepts low energy photons. Without dead regions, the quantum efficiencies of a simulation response at $\sim1\;{\rm keV}$ would greatly exceed those of the CALDB response. As shown in Figure \ref{fig:parameter_tuning_explanation} (a), we assumed the dead region is composed of two layers, ${\rm SiO_2}$ and Si. Throughout simulations for various sets of ${\rm SiO_2}$ and Si thicknesses, we determined the best parameter sets of the dead regions, as shown in the first and second rows of Table \ref{tab:parameter_tuning_result}. The FI CCDs yielded thicker dead regions, ${\rm SiO_2}\;0.6\;{\rm \mu m}+{\rm Si}\;0.4\;{\rm \mu m}$, while the BI CCD yielded a smaller dead region, ${\rm SiO_2}\;0.05\;{\rm \mu m}+{\rm no\;Si}$. These results are consistent with the existence of the readout structure on the surface of the photon illuminated side of the FI CCDs. It is also reasonable that the independent simulations of the two FI CCDs yielded exactly the same geometries of the dead regions.

The thickness of the depletion layer affects the transmission rates at high energies, in contrast to the dead regions. Although it is measured for each XIS by previous study \citep{Koyama2007}, our simulation displayed certain discrepancies with the CALDB responses in a high energy band (7--10 keV) when we adopted the reported values. Thus, we searched for the best value of the depletion layer thickness. They are shown in the third row of Table \ref{tab:parameter_tuning_result}. While the BI CCD yielded the same value as previously measured one ($42\;{\rm \mu m}$), the FI CCDs yielded thicker depletion layers, $70\;{\rm \mu m}$ for XIS0 and $74\;{\rm \mu m}$ for XIS3, than the previously reported value ($65\;{\rm \mu m}$).

\begin{figure*}
\begin{center}
\includegraphics[width=160mm]{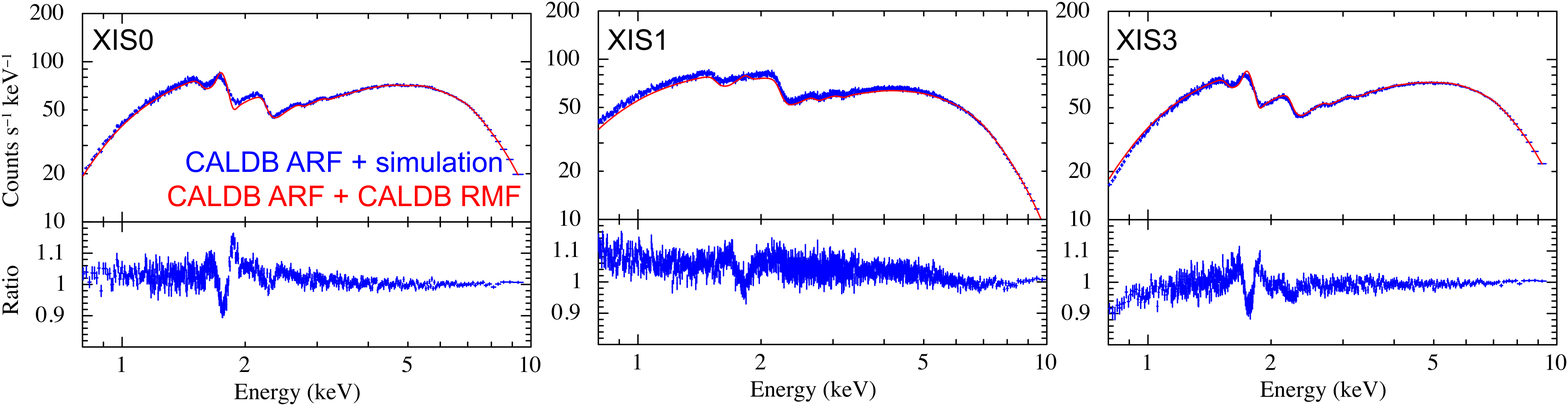}
\end{center}
\caption{Comparison of the responses generated by the simulation with those via the CALDB for XIS0, 1, and 3 in 0.8--10 keV. Red lines and blue crosses denote the CALDB responses and the simulation results, respectively. The error bars of the simulation results indicate $1\sigma$ statistical errors. The simulation responses were obtained at the best tuned parameters of the dead regions and the depletion layer thicknesses. We set $S(E)=S_0$ as the input source spectrum. Photon events within 1.7--1.9 keV were excluded in the parameter searches because the responses around the Si-K edge have large uncertainties.}
\label{fig:comp_sim_resp}
\end{figure*}

Figure \ref{fig:comp_sim_resp} shows the comparisons of count spectra between the simulations and the CALDB responses at the best-fit parameters of the dead regions and the depletion layer thicknesses. We excluded $<0.8\;{\rm keV}$ and $1.7$--$1.9\;{\rm keV}$ photons in parameter searches since the detector responses in these bands include large uncertainties due to constant tail components \citep{Koyama2007} and Si-K edge \citep{Okazaki2018}, respectively. The simulation agrees well with the CALDB response within a $\lesssim10\%$ accuracy for each detector. For the small differences left, we applied another small correction at the stage of ``event sampling'', which we describe in detail in Section \ref{section:effective_algorithm}.

\subsection{Parameter tuning with observation data}\label{section:parameter_tuning_2}
Events detected by an X-ray CCD are usually classified into various types of event shapes in order to distinguish whether they are originated from X-ray signals or from particle backgrounds. This classification (Grade categorization) was first introduced by \textit{ASCA} \citep{Tanaka1994} (also see the technical description of \textit{Suzaku} \footnote{\url{https://heasarc.gsfc.nasa.gov/docs/suzaku/analysis/abc/node1.html}}). The grade distribution, which is controlled by how much charge clouds diffuse, significantly affects the pile-up degree. The agreement with the CALDB response is not enough to simulate a spectrum affected by the pile-up because the CALDB response does not contain information about the grade distribution. Therefore, it is necessary to search for the best electric field structure, which directly controls the degree of charge diffusion, to reproduce the grade distribution. The reference has to be obtained from real observation data. In this step, we employed \textit{Suzaku} XIS observation data in which any pile-up effects are negligible for further parameter tuning.

\subsubsection{\textit{Suzaku} XIS observation}
We adopted an observation of PKS~2155-304 as a reference for XIS grade distribution. The observation log is presented in the first row of Table \ref{tab:observation_log}. Its source flux of $\sim0.8\;{\rm mCrab}$ with a 1/4 window mode observation corresponds to a full-window observation for an $\sim0.2\;{\rm mCrab}$ source, which is a sufficiently low flux for XIS to be free from the pile-up effects (also see Figure \ref{fig:model_simulation}). Note that $1\;{\rm Crab}$ denotes a flux of $2.4\times10^{-8}\;{\rm erg\;s^{-1}\;cm^{-2}}$ in the 2--10 keV band. After a data reduction by {\tt aepipeline}, for each CCD, we extracted a circular region with a radius of 120 pixels ($\sim\ang{;;125}$) from the image center, which covers $\sim90\%$ of the whole effective area \citep{Serlemitsos2007}. The XIS2 data were excluded since the detector was not in operations at this epoch. We then extracted spectra of three types of event shape which compose all the good event, single (Grade 0), double (Grade 2--4) and extended events (Grade 6) \footnote{The L-shape events and $2\times2$ square-shape events. For \textit{Hitomi} SXI and \textit{XRISM} Xtend-SXI, these events are classified into Grades 6 and 8.}.


\subsubsection{Determination of electric field structure}
The grade distribution is controlled by the electric field structure; stronger electric field leads to faster drift and less diffusion of charge clouds. As shown in Figure \ref{fig:parameter_tuning_explanation} (b), an ideal electric field structure in the depletion layer of a p-type CCD would be
\begin{eqnarray}
E(z)=E(l_{\rm dep})\frac{z}{l_{\rm dep}}=\left(\frac{V}{l_{\rm dep}}\right)\left(\frac{z}{l_{\rm dep}/2}\right),
\end{eqnarray}
where $l_{\rm dep}$ denotes the thickness of the depletion layer and its integral
\begin{eqnarray}
\int_0^{l_{\rm dep}}E(z)dz=V
\end{eqnarray}
gives the voltage applied between the two ends of the depletion layer. We treated the voltage $V$ as a simulation parameter, and searched for the best voltage that reproduces well the grade distribution in real observations.

The simulation was carried out with no pile-up effects by setting the frame exposure to a sufficiently short time, as mentioned in Section \ref{section:parameter_tuning_1}. For each CCD, we generated an unfolded spectrum of PKS~2155-304 from the observed spectrum by fitting it with a power-law model, and input it as an incident source spectrum for the simulation. The generated count spectra for various grades were compared with those derived from the real observation data.

\begin{figure*}
\begin{center}
\includegraphics[width=160mm]{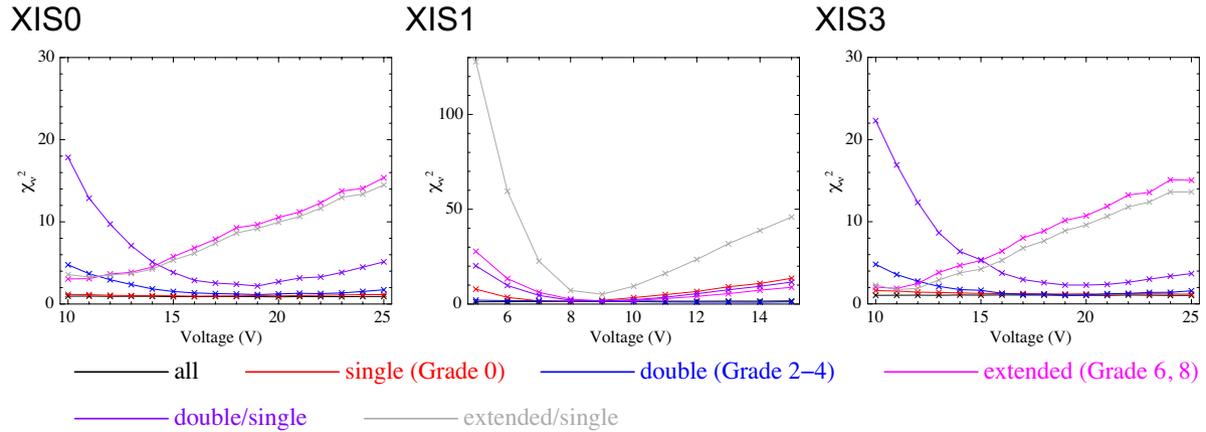}
\end{center}
\caption{\red{Reduced chi-square $\chi_{\nu}^2$} of the observation spectra against the corresponding simulations as functions of the simulation parameters of the bias voltages for XIS0, 1, and 3. We compare the simulation to the observation based on ratios between different event types as well as count spectra. Black, red, blue, and magenta lines represent count spectra of all, single, double, and extended events, respectively. Purple and gray lines denote double/single ratio and extended/single ratio, respectively.}
\label{fig:voltage_result}
\end{figure*}

Figure \ref{fig:voltage_result} shows the variations of observation-simulation discrepancies with the bias voltage. The most dominant grade, which is the single event for the FI CCDs (XIS0, 3) and the double event for the BI CCD (XIS1), keeps almost constant with voltage variation. Other grades, on the other hand, show large changes with the voltage. For the FI CCDs, we see a large difference between the value at the minimum $\chi_{\nu}^2$ for the double events and that for the extended events. This means that it is difficult to find the best parameter value for all event grades by varying only the voltage. Therefore, we decided to focus on optimizing the ratio of the double events to the single events, because the extended events populate only a few \% of all the events. As presented in Table \ref{tab:parameter_tuning_result}, we determined the best bias voltage to $19.0\;{\rm V}$ for XIS0 and $20.4\;{\rm V}$ for XIS3. For the BI CCD, although the difference between the best voltages for the single events and for the extended events are rather smaller, we correspondingly used the ratio of the double events to the single events. The result is shown in Table \ref{tab:parameter_tuning_result}; the best bias voltage was $9.0\;{\rm V}$ for XIS1.

\begin{figure*}
\begin{center}
\includegraphics[width=160mm]{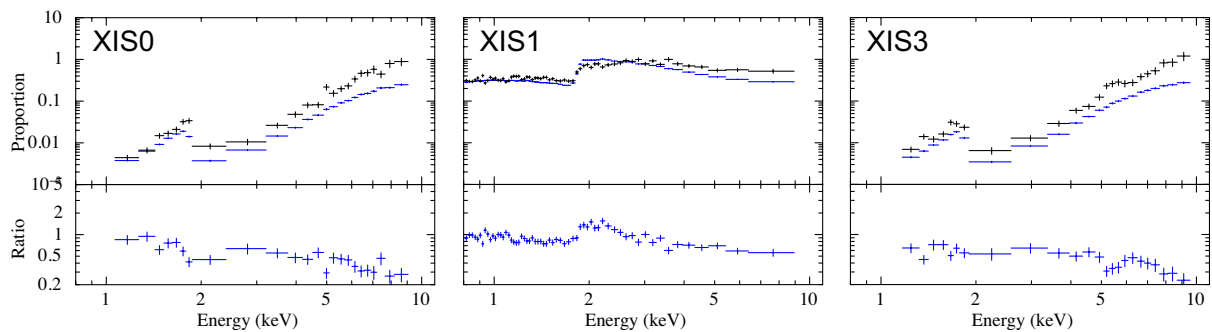}
\end{center}
\caption{Ratios between different event types as functions of photon energy for the PKS~2155-304 observation with \textit{Suzaku} XIS are plotted in black. Their corresponding simulations are shown in blue. The upper panels show ratios of double events (Grade 2--4) to single events (Grade 0), while the lower panels show ratios of extended events (Grade 6) to single events with energies. \redddd{The bottom plot of each panel shows the ratio between the observation and the simulation.} Photon events within 1.7--1.9 keV were excluded in the parameter searches because of large uncertainties in response. The error bars represent $1\sigma$ statistical errors.}
\label{fig:event_distribution}
\end{figure*}

Figure \ref{fig:event_distribution} shows comparisons between the observation data and the simulation for the grade distributions. The simulation successfully reproduced the double/single event ratio within an accuracy of $30\%$ for each detector. As we have discussed, the extended/single event ratios show large discrepancies between the observations and simulations at high energies (above $2\;{\rm keV}$), especially for the FI CCDs. However, the discrepancy in the extended events is negligible for the purpose of examining the pile-up effects, because the number of extended events in the high energy band is sufficiently small not to affect spectral parameters. In the case of the observation of PKS~2155-304 with XIS0, for example, the number of extended events of 2--10 keV is only $\sim0.9\%$ of all the 0.8--10 keV events, which means the systematic error by this discrepancy would be much less than $1\%$.

It is necessary to note that the best bias voltages obtained above do not match those in real observations. Still, the results are consistent with the fact that the voltages applied to the FI CCDs are higher than that applied to the BI CCD.

\section{Simulation of pile-up effects}\label{section:model_simulation}
In Section \ref{section:reproduction_of_observation}, we have obtained appropriate simulation parameters to reproduce \textit{Suzaku} XIS observations. In this section, we examine how a spectral shape would be changed or distorted as a result of the pile-up by using our simulator. We focused on pile-up effects on an absorbed power-law spectrum, which is one of the most typical and frequently used models in X-ray spectroscopy. The three parameters of the model---the hydrogen column density $N_{\rm H}$, the photon index, and the \redd{energy} flux (normalization)---characterize properties of celestial sources. Thus, it is essential to know how the pile-up would modify the values of these parameters when the data were analyzed via conventional methods and its dependence on the source brightness.

\begin{figure*}
    \begin{center}
    \includegraphics[width=160mm]{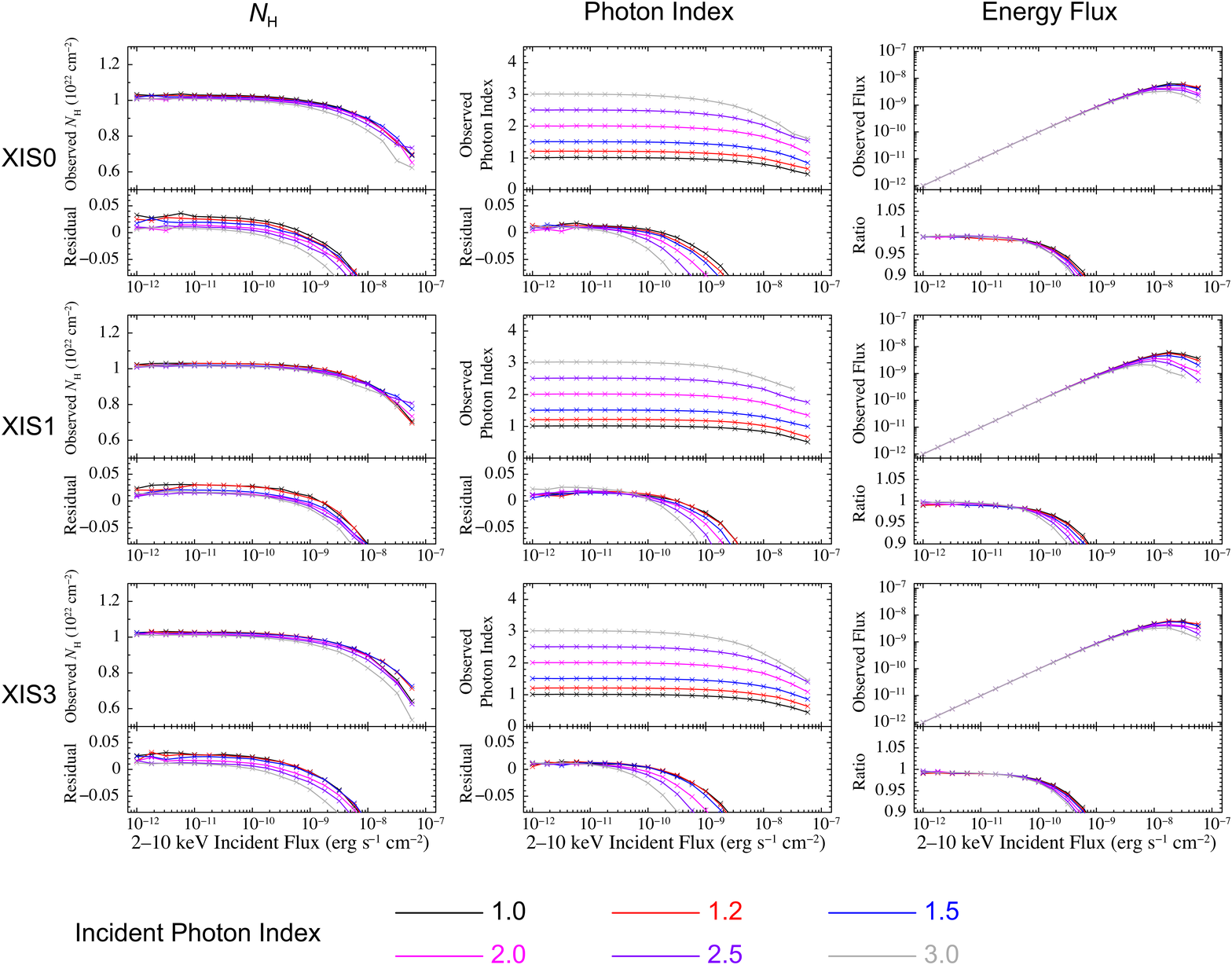}
    \end{center}
    \caption{The effects of the pile-up on \textit{Suzaku} XIS virtual observation data described by absorbed power-law spectra (0.8--10 keV). All the data were generated by simulations. The frame exposure was set to $8.0\;{\rm s}$, which is the default value of XIS. Each panel shows the fitting results (best-fit values) of a spectral parameter as a function of the incident source flux, and differences from the incident value by displaying residuals or ratios. The input $N_{\rm H}$ were fixed at $1.0\times10^{22}\;{\rm cm^{-2}}$ in all the cases. The first, second and third rows represent XIS0, XIS1, and XIS3, respectively. The first, second, and third columns represent the three parameters, $N_{\rm H}$, photon index, and 2--10 keV unabsorbed \redd{energy} flux, respectively. Each color corresponds to distinct photon index: $1.0$ (black), $1.2$ (red), $1.5$ (blue), $2.0$ (magenta), $2.5$ (purple), and $3.0$ (gray).}
    \label{fig:model_simulation}
\end{figure*}

\begin{figure*}
    \begin{center}
    \includegraphics[width=160mm]{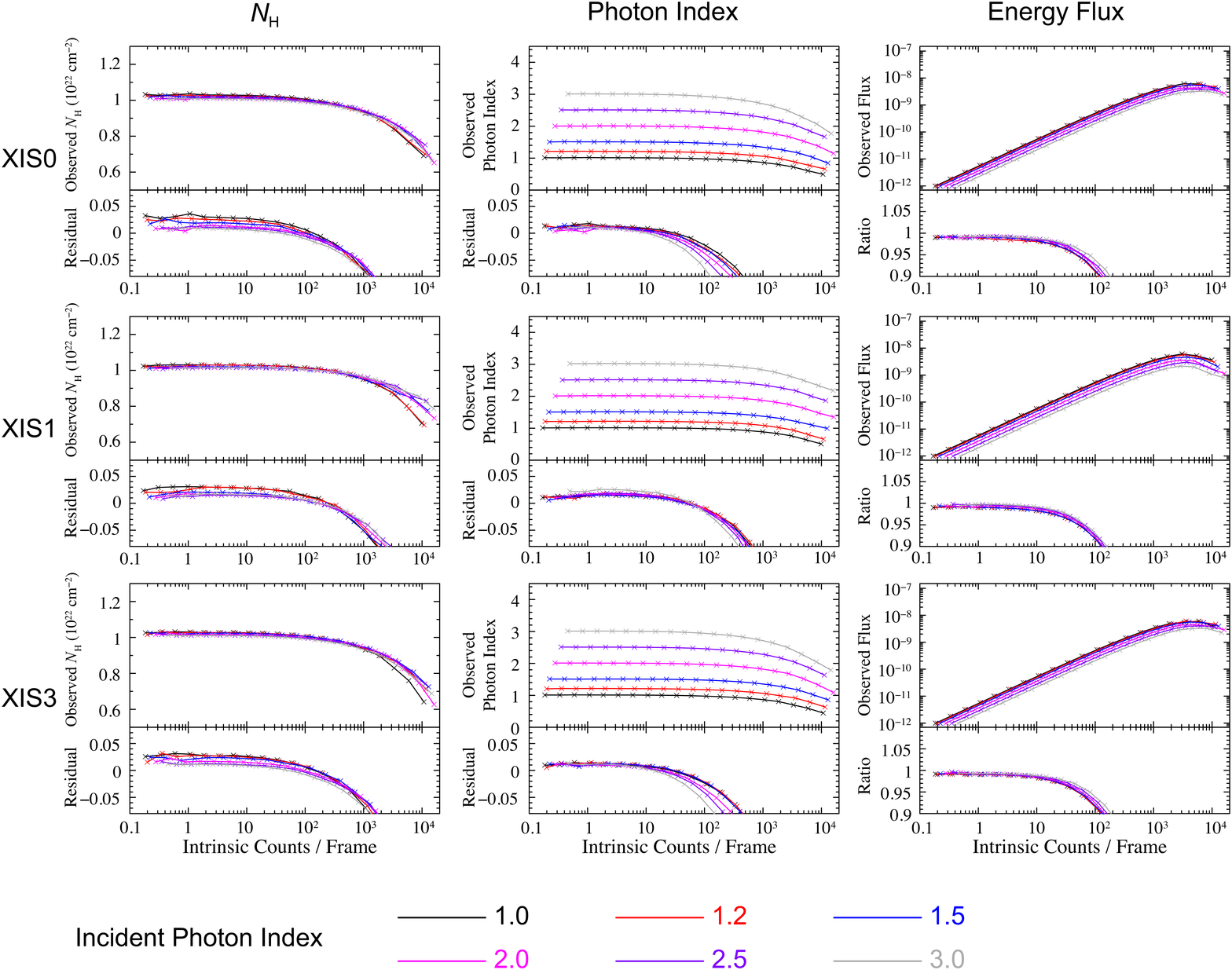}
    \end{center}
    \caption{\red{Same data as Figure \ref{fig:model_simulation} but plotted as a function of \reddd{intrinsic} counts per frame. The energy range of the count rate is 0.8--10.0 keV. The extraction area is limited within a radius of 120 pixels from the PSF center.}}
    \label{fig:model_simulation_cpf}
\end{figure*}

We performed simulations of absorbed power-law spectra with various photon indices and \redd{energy} fluxes with a constant $N_{\rm H}=1.0\times10^{22}\;{\rm cm^{-2}}$ \red{in an energy range of $0.2$--$12\;{\rm keV}$} to obtain the \reddd{observed} count spectrum for each case. We applied a conventional spectral analysis with {\tt XSPEC} to the simulated count spectrum. Then, we compared the obtained best-fit parameters with the input spectral parameters. Figure \ref{fig:model_simulation} shows how the obtained spectral parameters, $N_{\rm H}$, photon index, and \redd{energy} flux would change with increase of the source flux for the three detectors, XIS0, 1, and 3. \red{Figure \ref{fig:model_simulation_cpf} shows the same results but plotted as a function of \reddd{intrinsic} counts per frame.} 
Note that since the frame exposure is set to the XIS default value $8.0\;{\rm s}$, one needs to calculate the corresponding incident flux when the frame exposure is shorter. For example, if the observation is conducted using the 1/4 window mode and the frame exposure is $2.0\;{\rm s}$, the corresponding incident flux in Figure \ref{fig:model_simulation} would be 1/4 of the original value.

In all the panels of Figure \ref{fig:model_simulation}, we can see a simple tendency that while the parameters keep their original incident values in low-flux regions, they gradually move away from the incident values as the flux gets higher. This is clearly a consequence of the spectral distortion by the pile-up. The widely-known effects of the pile-up, such as hardening of photon index and decrease in flux, are well reproduced in our simulation. The hydrogen column density $N_{\rm H}$ is also decreased by the pile-up as if the absorption was lower, because it has a positive correlation with the photon index and \redd{energy} flux.

We also see a common trend of various pile-up degrees among different photon indices. Soft spectra start to show the pile-up effects in lower incident flux regions than harder spectra because the softer spectra contain more photons than harder ones with the same flux. For example, the XIS0 starts to show a photon index hardening at an incident flux of $\sim 10^{-11}\;{\rm erg\;s^{-1}\;cm^{-2}}$ for a photon index of $3.0$, while $\sim 10^{-10}\;{\rm erg\;s^{-1}\;cm^{-2}}$ for a photon index of $1.0$.
\red{Figure \ref{fig:model_simulation_cpf} shows that the pile-up effects depend less on photon indices when expressed as a function of \reddd{intrinsic} counts per frame. However, the spectral hardening still shows different behaviors among different photon indices; soft spectra start hardening at lower count rate. This is because low-energy photons tend to contribute more to spectral hardening than high-energy photons since the pile-up of the latter more likely produces bad events.}
\red{The same trend is shown in the previous pile-up study on {\it XMM-Newton} \citep{Jethwa2015}.}
Therefore, when evaluating the pile-up degree, one should take into account the contribution of the spectral shape as well as the incident flux.

The difference of the detector type also leads to different pile-up effects. The FI CCDs (XIS0, 3) show more spectral hardening than the BI CCD (XIS1) at the same incident spectrum, while they show less flux decline than the BI CCD. \red{In the case of a photon index of $3.0$ (gray lines), for example, the FI CCDs reach photon indices of $2.95$ at fluxes of $\sim1.5\times10^{-10}\;{\rm erg\;s^{-1}\;cm^{-2}}$, while the BI CCD reach the same point at a flux of $\sim5\times10^{-10}\;{\rm erg\;s^{-1}\;cm^{-2}}$.} This trend results from the difference in how piled-up photons are processed. For the FI CCDs, the single pixel events are the most dominant grade. Merging two single events due to pile-up will highly possibly form ``good events'' such as single or double pixel events. As a result, the pile-up effects would emerge as a spectral hardening rather than a flux decline. For the BI CCD, on the other hand, charge clouds get more easily diffused, and the double pixel events are the most dominant grade. Merging them will more likely form ``noise-like events'' extended to many pixels, which will be excluded from the \reddd{observed} count spectrum. Thus, the pile-up effects could emerge as a flux decline rather than a spectral hardening.

\section{Spectral analysis of pile-up affected data}\label{section:spectral_analysis_for_pile-up_data}
The main purpose of this work is to conduct spectral analysis of pile-up affected data without losing information that is contained in pile-up affected regions on the CCD. We picked up three sets of XIS observation data with different pile-up degrees: a pile-up-free (PKS~2155-304), a moderately pile-up affected (Aquila~X-1), and a strongly pile-up affected data set (Crab). We applied our framework and carried out spectral analyses on these observation data. The observation logs are presented in Table \ref{tab:observation_log}. For all the three data sets, the XIS2 data were not available since it was not under operations.

We adopt a forward method for spectral fitting, \redd{using our simulation to predict \reddd{observed} counts from a model}. We first assume a finite parameter space and divide it into discrete meshes. Then the simulator performs a full simulation for each point in the parameter space, generating the corresponding \reddd{observed} count spectrum. Finally, we evaluate the discrepancy with the observation data for each point and iterates this process until the best parameter set is found. The details of the best-fit parameter search and error estimation are described in Appendix.

A full simulation for a parameter set takes a long time ($\sim 1\;{\rm hour}$) to achieve sufficient statistics since the simulation of physical processes for each incident photon is so complicated. If we iterated full simulation runs for many points in the parameter space, it would take a few months to obtain the best parameter set for one observation with a Mac mini 2020 (Intel Core i7 (Gen 8), 6-core 3.2 GHz, and 10-thread parallel calculations). Therefore, we have constructed a new ``database sampling algorithm'', which significantly reduces time for the spectral fitting. Owing to the algorithm, the spectral analysis for pile-up affected data can be completed within a few days with our machine, and also the method realigns the spectra in the absence of pile-up with the CALDB prediction. In this section, we describe our database sampling algorithm and then present our results of spectral analyses on the \textit{Suzaku} XIS data.

\begin{table*}
  \tbl{List of \textit{Suzaku} XIS observations}{%
  \scalebox{0.9}{
  \begin{tabular}{cccccc}
      \hline
      Source & ObsID & \begin{tabular}{c}Date\\(YYYY-MM-DD)\end{tabular} & Observation Mode & \begin{tabular}{c}Frame exposure\\(s)\end{tabular} & \begin{tabular}{c}Net observation time\\(s)\end{tabular}\\ 
      \hline
      PKS~2155-304 & 101006010 & 2006-05-01 & 1/4 window & $2.0$ & $31473$\\
      Aquila~X-1 & 402053010 & 2007-09-28 & 1/4 window & $0.5$ & \begin{tabular}{c}$918$ (XIS0, 3)\footnotemark[$*$]\\$3456$ (XIS1)\end{tabular}\\
      Crab & 101004020 & 2006-04-04 & \begin{tabular}{c}1/4 window + burst (XIS0, 3)\\burst (XIS1)\end{tabular} & $0.1$ & \begin{tabular}{c}$957$ (XIS0, 3)\\$239$ (XIS1)\end{tabular}\\
      \hline
    \end{tabular}}}\label{tab:observation_log}
\begin{tabnote}
\footnotemark[$*$] The remaining $2538\;{\rm s}$ observations of XIS0 and 3 are conducted with a different event extraction algorithm ($2\times2$ mode).\\
\end{tabnote}
\end{table*}


\subsection{Database sampling algorithm}\label{section:effective_algorithm}
\begin{figure*}
    \begin{center}
    \includegraphics[width=160mm]{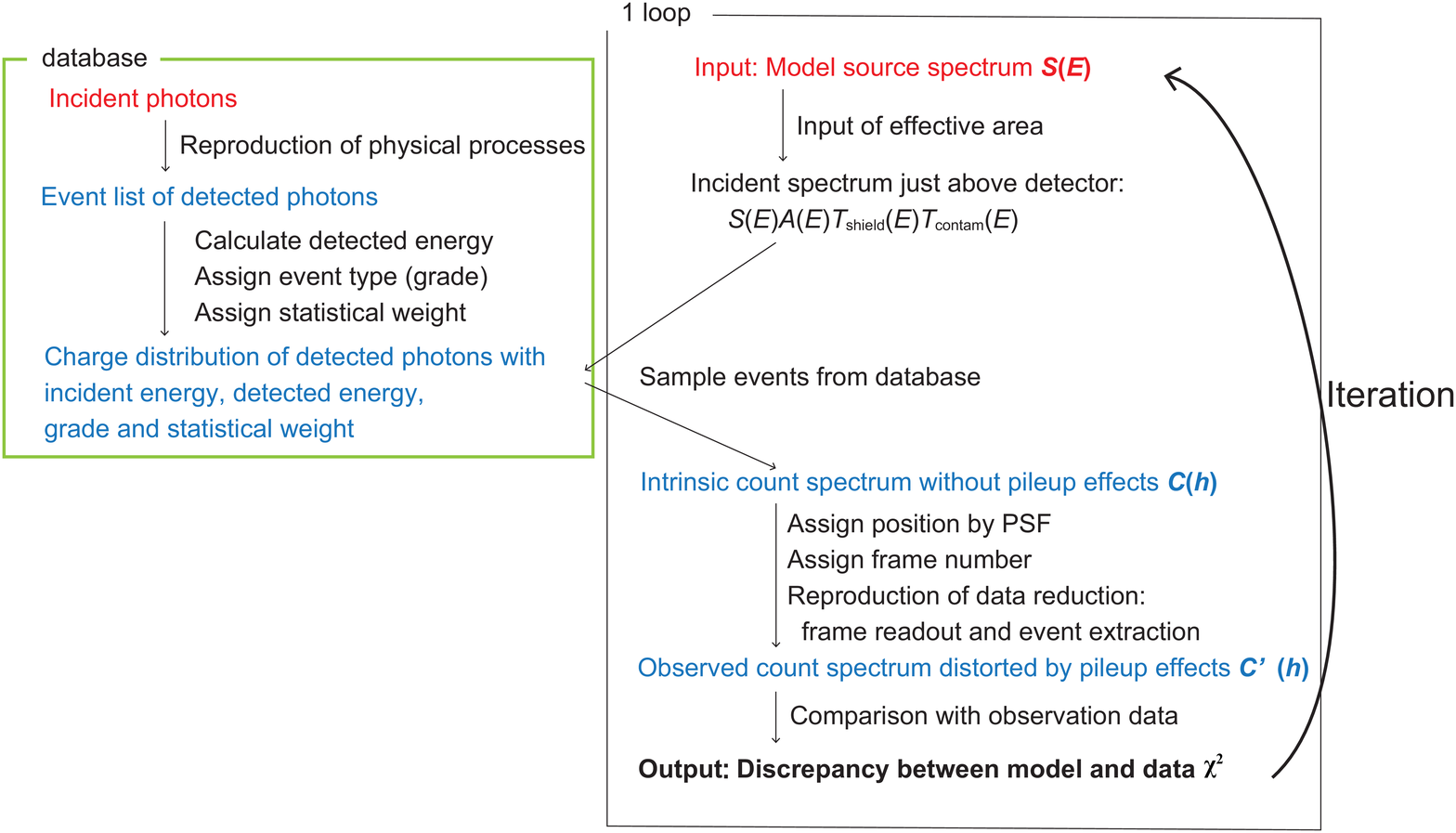}
    \end{center}
    \caption{Outline of our spectral analysis framework. The red and blue texts represent information of the photon spectrum and count spectrum, respectively.}
    \label{fig:spectral_analysis_algorithm}
\end{figure*}

Figure \ref{fig:spectral_analysis_algorithm} shows the outline of the spectral analysis framework using the database sampling algorithm. \red{In this method, complicated physical processes including the shield transmission, photoelectric absorption, charge drift and diffusion are pre-calculated for a sufficient number of incident photons with various incident energies. Their interactions with detectors are stored in the database, so that one can sample events from it and immediately reconstruct frame images when performing a spectral analysis. This method takes a much shorter time than simulating physical processes for every incident spectrum, but does not lose information about physical processes. In the following, we describe each step of the database sampling algorithm.}

\subsubsection{Generation of simulation database}
To generate the simulation database, we performed the full simulations for a sufficient number of incident photons with various energies; \redd{i.e., $2\times10^8\;{\rm photons}$ for each detector, sampled over a flat spectrum between $0$ and $12\;{\rm keV}$}. For each incident photon event, the event extraction algorithm is applied to the electrical signals generated by the detector. \redd{The information about charge distribution around an event is stored in the database (as a $5\times 5$ map of pixel charge following the telemetry format for Suzaku XIS) because it affects the pile-up.} In this process, whether the event is classified into an X-ray event or not is determined by the Grade algorithm, and the pulse height of the event is calculated. Therefore, we obtain a response matrix that can be compared with CALDB by relating detected pulse heights with incident photon energies as
\begin{eqnarray}
R_{\rm sim}(E_i, h_j)=\frac{N_{\rm sim}(E_i, h_j)}{\sum_{k} N_{\rm sim}(E_i, h_k)},
\end{eqnarray}
where we assume a discrete response matrix and $N_{\rm sim}(E_i, h_j)$ denotes the number of photons with an incident energy of $E_i$ and detected pulse height of $h_j$. Note that photons that produce ``Bad Grade'' events or penetrate the sensitive region with no interaction are all counted as $h=0$.

\subsubsection{Correction of simulation response}\label{section:correction_of_simulation_response}
The response matrix generated by the simulator shows slight discrepancies with the CALDB response. We correct the discrepancies by assigning a weight to each element of the simulation-generated response matrix. The weight we set for each event of the simulation database can be expressed as
\begin{eqnarray}
W(E_i, h_j)=W_0\frac{R_{\rm ref}(E_i, h_j)}{R_{\rm sim}(E_i, h_j)},
\label{eq:sampling_weight}
\end{eqnarray}
where the reference response matrix $R_{\rm ref}(E_i, h_j)$ denotes the CALDB response and $W_0$ represents the default sampling probability.
In this work, we set the default sampling probability to $W_0=0.5$ in order to meet $W(E_i, h_j)<=1$ at every $i$ and $j$.

\subsubsection{Sampling events from database}
The sampling of events from the database is based on an incident energy whose probability is given by the spectrum just above the detector. This spectrum is calculated by the multiplication of the incident model spectrum $S(E)$, effective area $A(E)$, and shield and contamination transmission rates $T_{\rm shiled}(E)$ and $T_{\rm contam}(E)$ (equation (\ref{eq:effective_area_all})). The analysis framework collects events from the database for all energy bins. The energies of these collected events are described by an \reddd{intrinsic} count spectrum $C(h)$ in Equation (\ref{eq:count_spectrum}), which is still free from pile-up effects. \redd{When the simulation picks up an event with an energy $E_{i}$ and a pulse height $h_{j}$, we draw an additional random number $w$ ($0<w<1$) and accept the event only if $w<=W(E_{i},h_{j})$ (see equation (\ref{eq:sampling_weight})).} Then, we simulate the CCD frame readout by assigning frame numbers to all the events and make frame images. To make an image, the positions of the events in a frame are sampled from the PSF. By re-extracting X-ray events from the frame images, we finally get the \reddd{observed} count spectrum which is affected by the pile-up effects $C'(h)$ (see Section \ref{section:development_of_CCD_simulator}).

\subsubsection{Benefits of database sampling algorithm}
\begin{figure}
    \begin{center}
    \includegraphics[width=80mm]{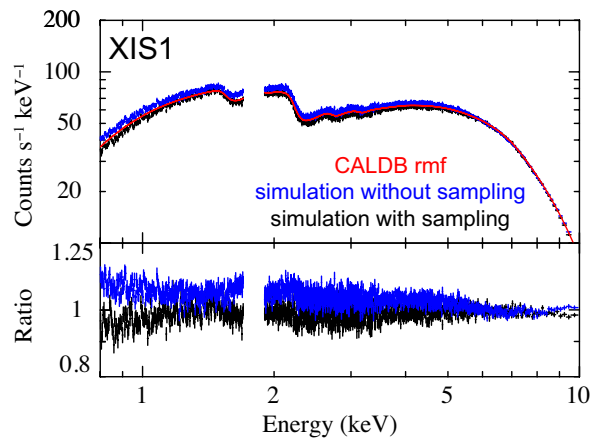}
    \end{center}
    \caption{Correction of simulation response by the database sampling algorithm. The red line and blue crosses denote CALDB response and simulation response without sampling, respectively, the same as Figure \ref{fig:comp_sim_resp}. The black crosses show the simulation response corrected through the database sampling algorithm. 1.7--1.9 keV photons are excluded due to large uncertainties in the response. The error bars of simulation results indicate $1\sigma$ statistical errors.}
    \label{fig:xis1_comp_sim_resp_sample}
\end{figure}

The database sampling algorithm improves our nonlinear spectral analysis in two aspects. One benefit is that it greatly reduces time taken for the end-to-end simulation to $\sim1/100$ while it keeps all information about charge signal distributions calculated by the full simulation. The other is that it reduces systematic errors of the simulator by the response correction. Figure \ref{fig:xis1_comp_sim_resp_sample} shows the correction of the simulation response for XIS1. While the quantum efficiencies of the simulation response without sampling slightly exceeds the CALDB response (also shown in Figure \ref{fig:comp_sim_resp}), the simulation response with sampling corrects this discrepancy, presenting a more accurate response matrix. \red{The reason why the agreement after correction is not perfect is probably due to the value of $W_0=0.5$; if the initial discrepancy exceeds a factor of $2.0$, namely $R_{\rm ref}(E_i, h_j)/R_{\rm sim}(E_i, h_j)>2.0$, the sampling probability of $W(E_i, h_j)$ will be underestimated.}

\subsection{Nonlinear spectral analysis applied to \textit{Suzaku} XIS data}\label{section:suzaku_analysis}

Incident spectra of all the three sources, PKS~2155-304, Aquila~X-1, and Crab, are well described by a single absorbed power law. The parameters of an absorbed power-law model are the hydrogen column density $N_{\rm H}$, photon index, and \redd{energy} flux (normalization). For all the observation data, we extracted X-ray events from a circular region with a radius of 120 pixels from the image center and performed two ways of spectral analyses designated as linear and nonlinear methods. The linear method denotes the conventional spectral fitting with {\tt XSPEC}, which does not take the pile-up effects into account. On the other hand, the nonlinear method uses our simulation-based framework that treats nonlinear effects caused by the pile-up in the spectral fitting. For the pile-up affected sources, we also present reference analyses of data that are not influenced by the pile-up either from regions of low count rates by applying the core-exclusion method or from other observatories with high timing resolutions. In all the spectral analyses, the energy bands with large response uncertainties are excluded, i.e., $<0.8\;{\rm keV}$, $>10\;{\rm keV}$, and $1.7$-$1.9\;{\rm keV}$. \red{Note that events in these excluded bands are taken into account in the pile-up simulations.} To reduce statistical errors of the simulations, the exposure times are set to \red{10 times as long as those of actual observations for PKS~2155-304, and 5 times for Aquila~X-1 and the Crab Nebula.}

\subsubsection{PKS 2155-304}\label{section:PKS2155-304}
The observation data of PKS~2155-304 we used here is the same as the one we used to optimize simulation parameters in section \ref{section:parameter_tuning_2}. The main purpose of applying our framework on this pile-up-free data is to confirm that the nonlinear method based on the simulation database has an equivalent capability of spectral fitting to the conventional method with {\tt XSPEC}. We compared the two ways of spectral analyses for 0.8--10 keV photons. The assumed spectral model is a pegged power-law ({\tt pegpwrlw}) without absorption, since the hydrogen column density of $N_{\rm H}\sim10^{20}\;{\rm cm^{-2}}$ \citep{Aharonian2009, Madsen2015cal} has a negligible impact above $0.8\;{\rm keV}$.

\begin{figure*}
    \begin{center}
    \includegraphics[width=160mm]{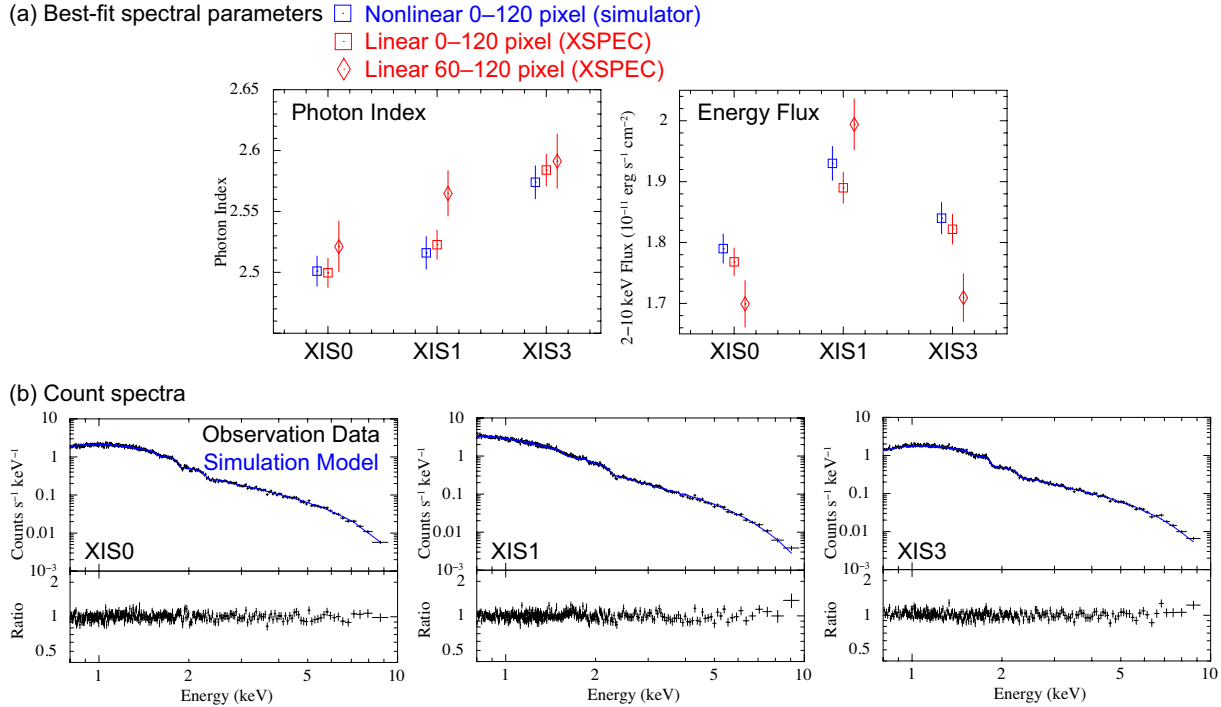}
    \end{center}
    \caption{Results of spectral analyses for PKS~2155-304 in 0.8--10 keV. (a) Best-fit parameters of the photon index and the \redd{energy} flux for XIS0, 1, and 3 derived from the three methods: our nonlinear spectral analyses using the entire 0--120 pixels (blue squares), the conventional linear spectral analyses with {\tt XSPEC} for the 0--120 pixel region (red squares), and the conventional analyses for the core-excluded 60--120 pixel region (red diamonds). Error bars represent 90\% confidence intervals. (b) \reddd{Observed} count spectra and fitting results of the nonlinear spectral analyses for XIS0, 1, and 3. Black crosses and blue lines denote the observation data and the best-fit model spectra, respectively. 1.7--1.9 keV photons are excluded in the parameter searches because of large uncertainties in response. Error bars represent $1\sigma$ statistical errors. The statistical errors of the simulations are not displayed in the upper panels, but are included in the error bars of the ratio plots in the lower panels.}
    \label{fig:PKS2155-304_fitting_result}
\end{figure*}

Figure \ref{fig:PKS2155-304_fitting_result} shows the results of spectral analyses with the two ways. Our framework successfully fitted observation data without any distinctive structures in residuals, and returned good $\chi_\nu^2$ (d.o.f.) values of $1.02\;(287)$, $1.16\;(298)$, and $1.05\;(241)$ for XIS0, 1, and 3, respectively. All of the best-fit parameters derived from the nonlinear spectral analyses are consistent with those yielded by the conventional linear spectral analyses with {\tt XSPEC} for the circular region (0--120 pixel), as shown in Figure \ref{fig:PKS2155-304_fitting_result} (a). Therefore, we confirm that our simulation framework has an equivalent capability of spectral analyses to the conventional manner at least for pile-up-free observation data. \red{Our method also yielded about twice smaller errors than those of the core-exclusion method for every best-fit parameter, which is due to the difference of statistics.} For the core-excluded region (60--120 pixel), \red{almost all the parameters agree with those yielded with the whole circular region within $2\sigma$. One exception is the flux of XIS3, in which the core-exclusion method underestimates the value. This could be due to uncertainties in the PSF distributions, but the confirmation needs additional analysis, which is beyond the scope of this work.}

\subsubsection{Aquila X-1}\label{section:AqlX-1}
Aquila~X-1 is a low mass X-ray binary composed of a neutron star and a main sequence star, which shows frequent fluctuations in its luminosity and spectral morphology \citep{Rutledge2002, Campana2014, Sakurai2014}. The \textit{Suzaku} observation we adopted here was carried out just after an outburst in 2007 September, showing an extremely bright luminosity compared to other periods \citep{Sakurai2014}. An observed flux of $\sim100\;{\rm mCrab}$ \citep{Sakurai2012} and a frame exposure of $0.5\;{\rm s}$ have a comparable impact of the pile-up as an observation for $\sim6\;{\rm mCrab}$ source with the full-window and the default exposure mode. This is expected to generate moderately pile-up affected observation data, which is evident from the condition with a flux of $\sim1.5\times10^{-10}\;{\rm erg\;s^{-1}\;cm^{-2}}$ and a photon index of $2.5$ in Figure \ref{fig:model_simulation}. Note that the data set was also examined by \citet{Yamada2012} as a moderately pile-up affected data set.

According to the study by \citet{Sakurai2012}, the wide-band spectra are well fitted by a ${\tt diskBB}+{\tt compPS}$ model, composed of a thermal component with a temperature of $kT\sim0.2\;{\rm keV}$ and a non-thermal component with a photon index of $\sim2.4$. Since the {\tt diskBB} and {\tt compPS} have many parameters, we adopted a simpler model instead, absorbed power law ({\tt phabs*pegpwrlw}), to make it easy to evaluate our framework. In order to eliminate the effects of discrepancy between this simple model and the data, we limited the energy band to 1--8 keV, excluding effects of the thermal component in the low energy band ($\lesssim 1\;{\rm keV}$) and the cut-off of the non-thermal component in the high energy band ($\gtrsim 8\;{\rm keV}$). For the conventional spectral analyses, we also performed spectral fitting for the pile-up-free region (an annulus with an inner radius of 60 pixels and outer radius of 120 pixels) as references.

\begin{figure*}
    \begin{center}
    \includegraphics[width=160mm]{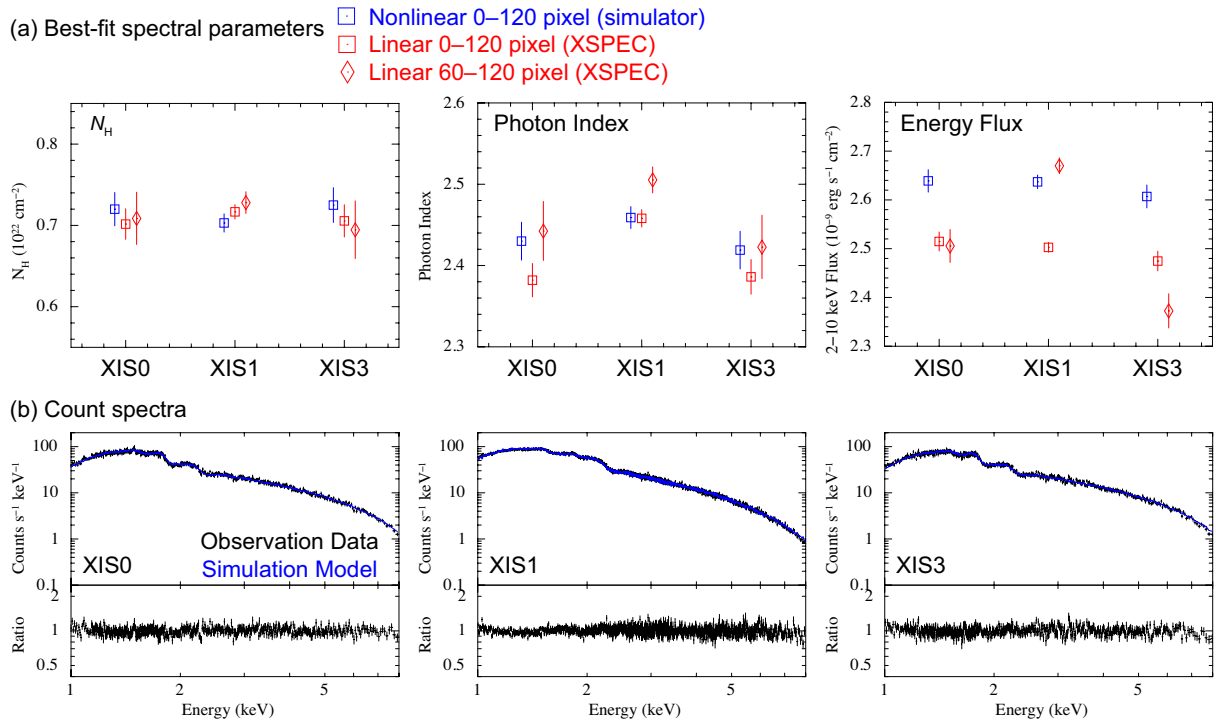}
    \end{center}
    \caption{Results of spectral analyses for Aquila~X-1 in 1--8 keV. (a) Best-fit parameters of the $N_{\rm H}$, the photon index, and the \redd{energy} flux for XIS0, 1, and 3 derived from the three methods: our nonlinear spectral analyses using the entire 0--120 pixels (blue squares), the conventional linear spectral analyses with {\rm XSPEC} for the pile-up affected 0--120 pixel regions (red squares), and the conventional analyses for the core-excluded 60--120 pixel regions (red diamonds). Error bars represent 90\% confidence intervals. (b) \reddd{Observed} count spectra and fitting results of the nonlinear spectral analyses for XIS0, 1, and 3. Black crosses and blue lines denote the observation data and the best-fit model spectra, respectively. 1.7--1.9 keV photons are excluded in the parameter searches because of large uncertainties in response. Error bars represent $1\sigma$ statistical errors. The statistical errors of the simulations are not displayed in the upper panels, but are included in the error bars of the ratio plots in the lower panels.}
    \label{fig:AqlX-1_fitting_result}
\end{figure*}

Figure \ref{fig:AqlX-1_fitting_result} shows the results of spectral analyses for Aquila~X-1. Our framework fitted well a model to the observation data without any distinctive structures in residuals, and returned good $\chi_{\nu}^2$ (d.o.f.) values of $1.01\;(454)$, $1.24\;(945)$, and $1.11\;(456)$ for XIS0, 1, and 3, respectively. \red{Our method also yielded about twice smaller errors than those of the core-exclusion method for every best-fit parameter, which is due to the difference of statistics.}
\redd{For both FI and BI CCDs, our framework showed photon indices consistent with the core-exclusion method.}
As for \redd{energy} fluxes, we see corrections of flux declines by the nonlinear method for all the CCDs, the blue-squared data points showing larger values than the red-squared points. The fluxes of the pile-up-free region disagree with the corrected results probably due to uncertainties of PSF distributions in the core-excluded region, \red{which is the same trend as Section \ref{section:PKS2155-304} and Figure \ref{fig:PKS2155-304_fitting_result} (a). To confirm it is due to the PSF problem, one needs additional analyses with good statistical precision, but at this point it is difficult to precisely evaluate the accuracy of our results for the \redd{energy} flux.} 

In brief, we confirm that the simulation-based framework is able to deal with moderately pile-up affected data and correct pile-up effects. Although the evaluation of the accuracy still remains ambiguous to some extent mainly due to the PSF uncertainties, it is remarkable that the nonlinear method yields best-fit spectral parameters highly consistent among the detectors for all the parameters. This suggests that the nonlinear method gives more reliable results than the core-exclusion method.

\subsubsection{Crab Nebula}\label{section:Crab}
The Crab Nebula, one of the brightest objects in X-ray astronomy, is a pulsar wind nebula which shows a power-law  spectrum with a photon index of $\sim2.1$ and an extremely high flux of $\sim2\times10^{-8}\;{\rm erg\;s^{-1}\;cm^{-2}}$ (2--10 keV) in the soft X-ray band. The \textit{Suzaku} data set obtained in a condition of a frame exposure of $0.1\;{\rm s}$ has a comparable impact of the pile-up as an observation of an $\sim12.5\;{\rm mCrab}$ source with a full window mode, which has a count rate twice that of the Aquila~X-1 data (section \ref{section:AqlX-1}) (also presented in \cite{Yamada2012}). Due to the strong pile-up, no previous research has successfully analyzed this set of observation data, while \citet{Kouzu2013} studied only the HXD data of the same data set. We performed spectral analyses on the observation data based on the absorbed power-law model in both linear and nonlinear ways. The reference to the pile-up-free region of these data is inappropriate, however, since the Crab Nebula has spatial variation that shows softer spectra with the distance from the central pulsar \citep{Madsen2015}. Although this effect is rather small with a spatial resolution of $\ang{;2.0;}$ with \textit{Suzaku} XRT \citep{Serlemitsos2007}, the outer regions still show \redd{higher fluxes than expected for a point source} and softer photon indices than the inner regions, which is consistent with the radial profile of Crab Nebula spectra investigated by a previous study with \textit{NuSTAR} \citep{Madsen2015}. Instead of the pile-up-free region, we used data obtained with other observatories with higher timing resolutions as references of spectral parameters, including \textit{Suzaku} HXD \citep{Takahashi2007}, \textit{NuSTAR} \citep{Harrison2013}, and \textit{Hitomi} SXS \citep{Mitsuda2014}.

\begin{figure*}
    \begin{center}
    \includegraphics[width=160mm]{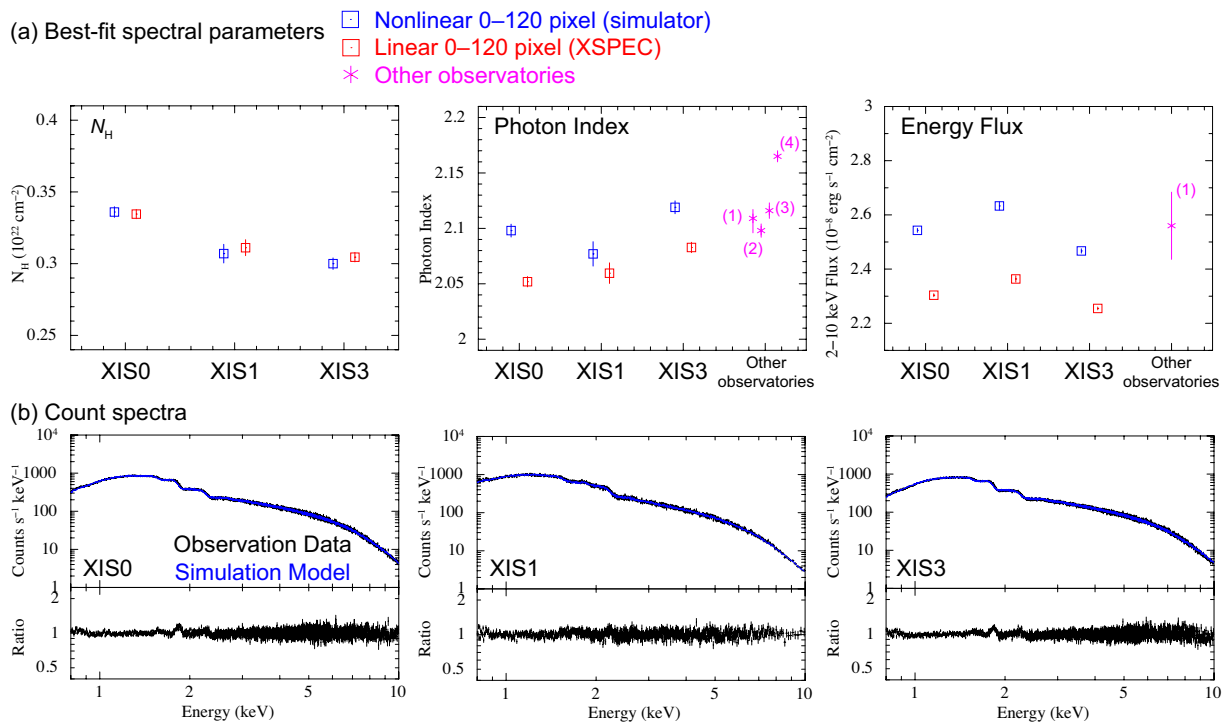}
    \end{center}
    \caption{Results of spectral analyses for the Crab Nebula in 0.8--10 keV. (a) Best-fit parameters of the $N_{\rm H}$, the photon index, and the \redd{energy} flux for XIS0, 1, and 3 derived from the two methods: our nonlinear spectral analyses using the entire 0--120 pixels (blue squares), the conventional linear spectral analyses with {\rm XSPEC} for the pile-up affected 0--120 pixel regions (red squares). Magenta data points denote the results of other observatories; (1) \textit{Suzaku} HXD in 2006 \citep{Kouzu2013}, (2) \textit{NuSTAR} in 2015 \citep{Madsen2017}, (3) \textit{NuSTAR} in 2016 \citep{Madsen2017}, (4) \textit{Hitomi} SXS in 2016 \citep{HitomiCollaboration2018Crab}. Error bars represent 90\% confidence intervals. (b) \reddd{Observed} count spectra and fitting results of the nonlinear spectral analyses for XIS0, 1, and 3. Black crosses and blue lines denote the observation data and the best-fit model spectra, respectively. 1.7--1.9 keV photons are excluded in the parameter searches because of large uncertainties in response. Error bars represent $1\sigma$ statistical errors. The statistical errors of the simulations are not displayed in the upper panels but are included in the error bars of the ratio plots in the lower panels.}
    \label{fig:Crab_fitting_result}
\end{figure*}

Figure \ref{fig:Crab_fitting_result} shows the results of spectral analyses for the Crab Nebula. The models fitted by the nonlinear method agree well with the observation data without any distinctive structures in residuals, and returned good $\chi_{\nu}^2$ (d.o.f.) values of $1.28\;(1660)$, $1.22\;(847)$, and $1.31\;(1657)$ for XIS0, 1, and 3, respectively. The main effects of the pile-up, hardening of photon indices and decline of fluxes, are well corrected by the nonlinear spectral analyses, while the $N_{\rm H}$ were invariant between linear and nonlinear spectral analyses. The photon indices derived from the nonlinear spectral analyses presented consistent values with those derived from \textit{Suzaku} HXD and \textit{NuSTAR}, although they were harder than that yielded by \textit{Hitomi} SXS. \red{This could be due to the systematic uncertainties of SXS, which tends to obtain softer photon indices for power-law spectra \citep{Tsujimoto2018}.}
For the \redd{energy} flux, the nonlinear analyses also yielded consistent values with the result of the simultaneous \textit{Suzaku} HXD observation. The \redd{energy} fluxes yielded by \textit{NuSTAR} and \textit{Hitomi} SXS 10 years after the \textit{Suzaku} observation are much lower than our results since the Crab Nebula experiences a decrease in X-ray flux \citep{Wilson-Hodge2011}. Therefore, our nonlinear spectral analyses using the simulator yielded more reasonable spectral parameters than the conventional linear method with the core exclusion.


\section{Discussion}\label{section:discussion}
The conventional approach for pile-up affected data of CCDs has employed the core-exclusion method. Although this method has been effective in many previous researches, both statistical and systematic errors become large due to the loss of the events within the image center. The exclusion of the image core also produces uncertainties of the effective area since it depends on the PSF and the history of the satellite attitude fluctuation. This makes the flux estimation more difficult. The framework we proposed, on the other hand, suppresses both of the uncertainties by using the photons in the entire region including the pile-up affected regions, greatly improving the accuracy of spectral parameter estimations. Our method is beneficial particularly for time-variable sources since the improvement allows us to define finer time bins in light curves to trace temporal nature of the sources.
The framework can be universally applied to any observation data of photon-counting CCD once the simulation parameters are appropriately optimized. It will be useful for future X-ray missions such as \textit{XRISM} \citep{Tashiro2018} as the quantitative estimation of the pile-up effects is important for observation planning as well as development of spectral analysis methods.
In this section, we discuss limitations of the framework and a new useful indicator for evaluating a pile-up degree. 

\subsection{Limitation of the nonlinear fitting method}
As described in Section \ref{section:reproduction_of_observation}, the simulation after the parameter tuning is in good agreement with the CALDB detector response. The small discrepancies which still remain can be corrected by the data sampling algorithm introduced in Section \ref{section:correction_of_simulation_response}. However, the simulator is not able to completely reproduce grade distributions, which are not contained in CALDB and can only be obtained from real measurements. In fact, for the \textit{Suzaku} data, the simulation underestimates ratios of the extended events when the double/single event ratios are optimized to the data. This underestimation is probably due to the simple assumption of electric field structure in a CCD; a horizontally uniform field distribution is assumed. In reality, electric field near pixel boundaries may be weaker than that around a pixel center, producing more extended events. It might be necessary to consider a more realistic horizontal distribution of the electric field for a more accurate detector response.


For extremely pile-up affected data, even the nonlinear method may not work since too many photons would be lost by the pile-up. We need to define an upper limit of the incident flux for a safe application of the method. For a power-law spectrum with a typical value of $N_{\rm H}\sim1.0\times10^{22}\;{\rm cm^{-2}}$ observed by \textit{Suzaku} XIS, the maximum incident flux to safely apply the method would be the peak values of the observed flux in Figure \ref{fig:model_simulation} (right), which would be $\sim 1\;{\rm Crab}$ or $\sim10^{-8}\;{\rm erg\;s^{-1}\;cm^{-2}}$ in 2--10 keV. If one applies the nonlinear fitting method above the threshold, the observed flux becomes a bivalent function of the incident flux, which may lead to an ambiguous estimation of the incident flux. \red{This could be avoided by adopting a larger extraction region than that with a radius of 120 pixels, since it is shown that the observed count rate of a point source integrated over an infinite extraction region never decreases (see Section 3.2 of \citet{Ballet1999}.)}

\subsection{A new indicator for pile-up degree}
The pile-up fraction was suggested by \cite{Yamada2012} (also following \cite{Davis2001}) for an indicator of the degree of the pile-up. This is expressed as
\begin{eqnarray}
f_{\rm pl}(x)=\frac{\sum_{k=2}^{\infty}P(k, x)}{\sum_{k=1}^{\infty}P(k, x)},
\end{eqnarray}
where $x$ is the mean counts per frame per pixel and $P(k, x)$ is the Poisson distribution function with an expectation value of $x$. Thus, $f_{\rm pl}$ is interpreted as the ratio of the number of piled up photons to the number of incident photons. The condition assumed here is a uniform distribution of incident photons and no charge distributions among multiple pixels. In real observations, however, spatial distributions of incident photons from bright point sources follow the PSF. Multi-pixel events form a significant part of all the events and their fraction even varies with incident energies. Therefore, $f_{\rm pl}$ can be properly used under limited conditions.

We propose another indicator of the degree of the pile-up after the previously introduced $f_{\rm pl}$, taking into account the spectral morphology, PSF distributions, and charge distributions. This quantity is defined as the fraction of lost photons by the pile-up:
\begin{eqnarray}
F_{\rm pl}\left(S(E)\right)=1-\frac{\int_{h_{\rm min}}^{h_{\rm max}}C'(h)dh}{\int_{h_{\rm min}}^{h_{\rm max}}C(h)dh},
\end{eqnarray}
where $C(h)$ and $C'(h)$ denote the \reddd{intrinsic} count spectrum without the pile-up effects and the \reddd{observed} count spectrum distorted by the pile-up effects, respectively. The simulation framework can easily calculate this value by running one set of simulations.

\begin{figure}
    \begin{center}
    \includegraphics[width=80mm]{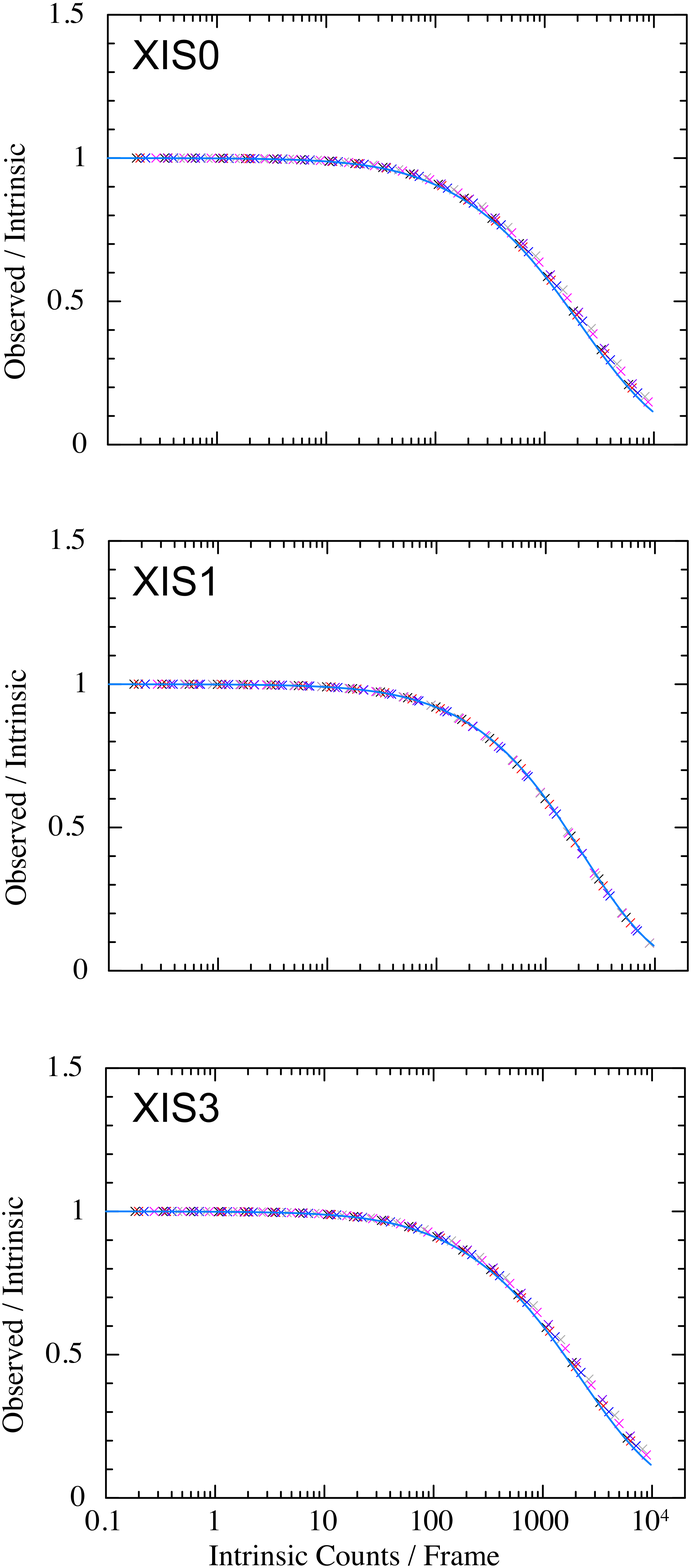}
    \end{center}
    \caption{\reddd{\redddd{Ratios} of observed photon counts to intrinsic photon counts ($1-F_{\rm pl}(S(E))$) as functions of intrinsic counts per frame for all the valid events in 0.8--10 keV for XIS0, XIS1, and XIS3 (x-mark). The incident spectral shape is assumed to be an absorbed power law with $N_{\rm H}=1.0\times10^{22}\;{\rm cm^{-2}}$. Different colors denote different photon indices, as presented in Figure \ref{fig:model_simulation}. The blue lines denote analytical calculations \redddd{of the ratios between observed and intrinsic single event counts} based on \citet{Ballet1999}.}}
    \label{fig:lost_photons}
\end{figure}

\begin{figure}
    \begin{center}
    \includegraphics[width=80mm]{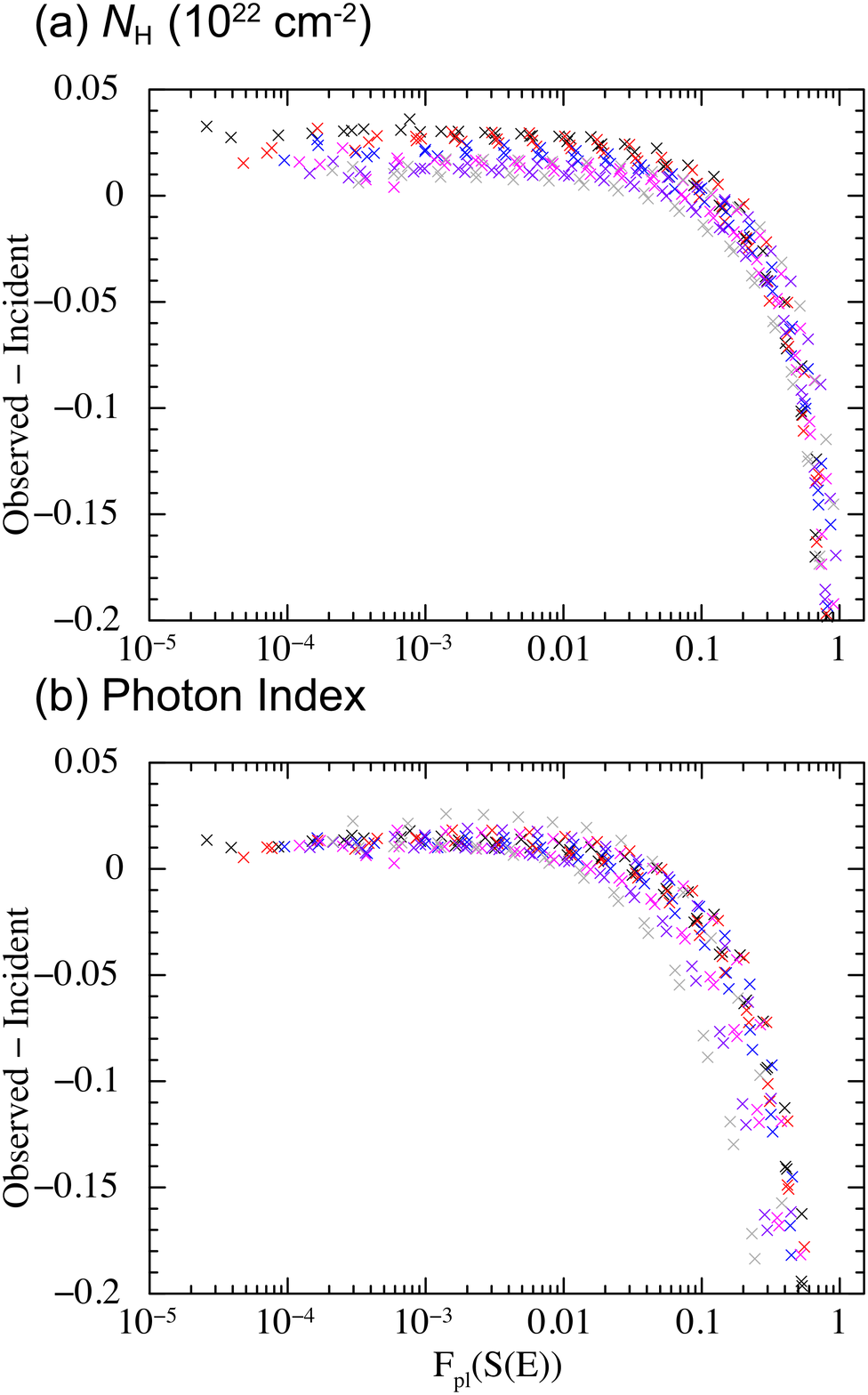}
    \end{center}
    \caption{Errors of spectral parameters as functions of the lost photon fraction $F_{\rm pl}(S(E))$. Simulation results for XIS0, 1, and 3 are plotted together. The assumed spectra are absorbed power-law with $N_{\rm H}=1.0\times10^{22}\;{\rm cm^{-2}}$, same as Figure \ref{fig:lost_photons}. Different colors denote different photon indices, as presented in Figure \ref{fig:model_simulation}. }
    \label{fig:pile-up_deg_parameter}
\end{figure}

Figure \ref{fig:lost_photons} presents one example of \reddd{$1-F_{\rm pl}\left(S(E)\right)$} for XIS0, 1, and 3 \redd{as a function of \reddd{intrinsic} counts per frame}. The assumed spectral shape is an absorbed power-law with $N_{\rm H}=1.0\times10^{22}\;{\rm cm^{-2}}$ and several photon indices. The fraction of \reddd{observed photons to intrinsic photons decreases} monotonically with the incident flux, as one can easily expect. \redd{The simulation results are in good agreement with a simplified analytical approach based on \citet{Ballet1999}, which calculates the lost photon ratio of single pixel events from the grade distribution and the PSF (see \redddd{equation (2)} and Figure 3 of \citet{Ballet1999}).}
Figure \ref{fig:pile-up_deg_parameter} shows the relation between $F_{\rm pl}(S(E))$ and the errors of spectral parameters obtained by the conventional linear method. The spectral shapes become slightly distorted at $F_{\rm pl}(S(E))\sim1\%$ and heavily distorted at $F_{\rm pl}(S(E))\sim10\%$. Combining Figures \ref{fig:lost_photons} and \ref{fig:pile-up_deg_parameter}, we would set benchmark values of an incident flux at $\sim1\;{\rm mCrab}\;(\sim2\times10^{-11}\;{\rm erg\;s^{-1}\;cm^{-2}})$ for the start of moderate pile-up, and $\sim10\;{\rm mCrab}$ for the start of strong pile-up. A similar estimation for a future CCD detector such as XRISM-Xtend would be useful for observation planning.



\section{Conclusions}\label{section:conclusions}
We constructed a simulation-based method of spectral analysis which can be applied to pile-up affected observation data of X-ray CCDs. This simulation treats physical processes of photons with a detector, charge transport in the detector, signal generations in pixels, frame readout of a CCD, and data reduction processes carried out on the satellite system and on the ground. We also developed a spectral fitting framework that considers a nonlinear detector response for pile-up affected data. This fitting method uses a database of X-ray events which is pre-computed by the simulator in order to reduce the computation time taken by the full simulations.

The framework was applied to \textit{Suzaku} XIS observations for its validation using real observation data. The detector response, quantum efficiencies, and event grade distributions of \textit{Suzaku} XIS were successfully reproduced after optimizing the simulation parameters. We then evaluated spectral distortion in the case of an absorbed power-law model, and obtained spectral parameter changes for different detector types and spectral properties. 
Finally, we performed spectral analysis of \textit{Suzaku} XIS data by using the framework, yielding reasonable and consistent results for three sources at different levels of the pile-up: PKS~2155-304 (negligible pile-up), Aquila~X-1 (moderate pile-up), and the Crab Nebula (strong pile-up). For the pile-up affected data, we successfully corrected the spectral hardening and the flux decrease which would appear in the conventional method.

Our framework based on Monte Carlo simulation solves pile-up problems without any loss of information from the image core of a bright point-like source. The method in this work can be applied to all types of X-ray CCDs by appropriate optimization of the simulation parameters.


\begin{ack}
The anonymous referee improved this work with his or her valuable comments.
\red{We acknowledge the support all the JAXA members who have contributed to the XRISM project, especially the members of XRISM MOPT group.}
We thank Tadayuki Takahashi for his suggestions and advice especially about the spectral analysis method.
We thank Hideki Uchiyama, Takanori Yoshikoshi, Hiromasa Suzuki, and Atsushi Tanimoto for valuable advice which improved our study. This work is supported by the Japan Society for the Promotion of Science (JSPS) Research Fellowship for Young Scientist No. 20J20050 (TT), JSPS/MEXT KAKENHI grant numbers 18H05861, 19H01906, 19H05185 (HO), 18H05459, 19K03908 (AB), and 16H03983 (KM). This work is also supported in part by Shiseido Female Researcher Science Grant (AB).

\end{ack}

\appendix 

\section*{Best-fit parameter search and error estimation}\label{section:parameter_search}
\subsection{Parameter search}
In the nonlinear spectral analysis, we search for the best parameter set by dividing the parameter space into discrete points. When dealing with $m$ parameters, $p_1,\;p_2,\;...p_m$, we divide the parameter space of $m$ dimensions into grids. If the i-th parameter axis is divided into $n_i$ points like ${p_{i1},\;p_{i2},...p_{in_i}}$, the total number of iteration for the parameter search will be
\begin{eqnarray}
N=\prod_{i=1}^{m} n_i.
\end{eqnarray}
Investigating all the points, the one showing the minimum $\chi^2$ can be determined as the best parameter set. However, due to \red{statistical} errors, our spectral analysis often presents some local minimum points. Therefore, we apply a function smoothing technique on the results like
\begin{eqnarray}
&\chi^2_{\rm smooth}&(p_{1x_1},...,\;p_{mx_m})\nonumber\\
&=&\frac{1}{3^m}\left(\sum_{y_1=x_1-1}^{x_1+1}...\sum_{y_m=x_m-1}^{x_m+1}\chi^2(p_{1y_1},...\;,p_{my_m})\right),
\end{eqnarray}
where we take the average of the $\chi^2$ values for all the surrounding points. This process is iterated until local minima vanish. As a result, we can determine the best parameter set taking the minimum $\chi^2$.

If we divide the parameter space in too fine steps, the time it takes for the spectral analysis will be too long. \redd{Therefore, for each spectral analysis, we start with a rough grid and perform the grid search three or four times, each time refining the grid step and narrowing the scope of the parameter space around the best fit.}

\subsection{Error estimation}
In the conventional spectral analysis with {\tt XSPEC}, the error estimation is done by investigating the variation of $\chi^2$ from the minimum point. For example, such points that presents $\Delta\chi^2<2.706$ are treated to be within 90\% confidence level. The correlation between parameters are also taken into account.

In our spectral analyses, the error estimation including parameter correlations is not realistic because it requires large amount of parameter searches. However, the error for a single parameter, excluding correlations with other parameters, is rather easy to calculate. By investigating the relation between $\chi^2$ and the parameter value around the minimum point and fitting it by a quadratic curve, we can estimate the error range. If the fitted quadratic curve is $\chi^2=ap^2+bp+c$, where $p$ is the parameter value, the range of 90\% confidence level for single parameter will be
\begin{eqnarray}
\Delta p=2\sqrt{2.706/a}.
\end{eqnarray}

Since we can get only the single parameter error without effects of correlations, it is necessary to estimate the real error from it. We assume that the degree of parameter correlation is the same for the linear and nonlinear spectral analysis, and calculate the error as
\begin{eqnarray}
\Delta p_{\rm nonlinear}=\Delta p_{\rm nonlinear,single}\times\frac{\Delta p_{\rm linear}}{\Delta p_{\rm linear, single}},
\end{eqnarray}
where all the three values of the right hand can easily be derived. In all the spectral analyses, $\Delta p_{\rm nonlinear}$ and $\Delta p_{\rm linear}$ present equivalent values to each other, which shows our approach for error estimation is reasonable.

\bibliography{pileup.bib}

\end{document}